\documentclass[review]{elsarticle}

\journal{Computer-Aided Civil and Infrastructure Engineering}

\usepackage{graphicx}
\usepackage{latexsym}
\usepackage{color}
\usepackage{caption}
\usepackage{subcaption}
\usepackage{bm}
\usepackage{bbm}
\usepackage{amsmath,amssymb, amsthm}
\usepackage{algorithm}
\usepackage{algpseudocode}
\usepackage{color,soul}
\usepackage{graphics}
\usepackage{amsfonts}
\usepackage{setspace}   
\usepackage{wasysym}
\usepackage{multirow}

\usepackage{adjustbox}
\usepackage{graphicx}
\usepackage{multirow}
\usepackage{nomencl}
\usepackage{float}
\usepackage[table]{xcolor}
\usepackage{standalone}
\usepackage{tikz}
\makenomenclature
\biboptions{sort&compress}

\newcommand{\bigO}{\ensuremath{\mathcal{O}}}

\makeatletter
\newcommand{\thickhline}{%
	\noalign {\ifnum 0=`}\fi \hrule height 1pt
	\futurelet \reserved@a \@xhline
}
\newcolumntype{"}{@{\hskip\tabcolsep\vrule width 1pt\hskip\tabcolsep}}
\makeatother

\newcolumntype{M}[1]{>{\centering\arraybackslash}m{#1}}

\usepackage{geometry}
\begin{document}

\begin{frontmatter}

\title{Deep Learning for Accelerated Reliability Analysis of Infrastructure~Networks}

\author{Mohammad Amin Nabian}
\author{Hadi Meidani\corref{mycorrespondingauthor}}
\address{Department of Civil and Environmental Engineering, University of Illinois at Urbana-Champaign, Urbana, Illinois, USA.}
\cortext[mycorrespondingauthor]{Corresponding author}
\ead{meidani@illinois.edu}

\begin{abstract}
Natural disasters can have catastrophic impacts on the functionality of infrastructure systems and cause severe physical and socio-economic losses. Given budget constraints, it is crucial to optimize decisions regarding mitigation, preparedness, response, and recovery practices for these systems. This requires  accurate and efficient means to evaluate  the infrastructure system reliability. While numerous research efforts have addressed and quantified the impact of natural disasters on infrastructure systems, typically using the Monte Carlo approach, they still suffer from high computational cost and, thus, are of limited applicability to large systems. This paper presents a deep learning framework for accelerating  infrastructure system reliability analysis. In particular, two distinct deep neural network surrogates are constructed and studied: (1) A \emph{classifier} surrogate which speeds up the connectivity determination of networks, and (2) An \emph{end-to-end} surrogate that replaces a number of components such as roadway status realization, connectivity determination, and connectivity averaging. The proposed approach is applied to a simulation-based study of the two-terminal connectivity of a California transportation network subject to extreme probabilistic earthquake events. Numerical results highlight the effectiveness of the proposed approach in accelerating the transportation system two-terminal reliability analysis with extremely high prediction accuracy.
\end{abstract}


\end{frontmatter}


\section{Introduction} \label{sec:introduction}

The hazard reliability for an infrastructure system is defined to be the degree of assurance that the system will continue to successfully operate at a desired level of performance during a certain period of time and in a specified environment in the aftermath of a hazard \cite{zacks2012introduction}. Assessment of the impact of natural disasters on infrastructure systems is of importance toward four main objectives: (1) Planning for actions that  eliminate or reduce the long-term risk to human life and infrastructure systems (e.g.\cite{godschalk1999natural}); (2) Disaster preparation or adjustment, which aims to reduce the risk of damages and injuries while enabling the capability to cope with the temporary disruption of the infrastructure systems (e.g.\cite{paton2003disaster}); (3) Development of effective emergency response strategies (e.g.\cite{perry2007natural}); and (4) Post-disaster recovery planning  (e.g.\cite{adie2001holistic}). These four are, respectively, known as the mitigation, preparedness, response, and recovery practices.

A variety of analytical \cite{wang2008integrated}, simulation \cite{stern2017accelerated,bocchini2011stochastic,chang2010transportations,bocchini2011generalized,nabian2017uncertainty}, and optimization  \cite{liu2009two} approaches are proposed in the literature for hazard reliability analysis of infrastructure systems. A comprehensive literature review on transportation infrastructure system performance in disasters is provided in \cite{faturechi2014measuring}. Simulation-based reliability assessment of large  infrastructure systems are often computationally intractable or expensive due to the large number of network components, complex network topology, statistical dependence between component failures, and uncertainties in the hazard models. This will impose limitations on design optimization or sensitivity analysis of these systems.  Alternatively, a more efficient response assessment  for large infrastructure systems can be made possible by using approximate surrogates \cite{koziel2013surrogate}.

Surrogates are fast  models that approximately describe the relationship between the system inputs and outputs and serve as a substitute for more expensive simulation tools. If the response evaluated by the reference expensive model is denoted  by $f\left ( \mathbf{x} \right )$,  a surrgate seeks to provide a global approximate function $\tilde{f}\left ( \mathbf{x} \right )$. This is typically done by using  a set of inputs $\mathbf{x}_{i} \in D^{d}, i=\left \{ 1,2,...,M \right \}$, and the corresponding `exact' system outputs $f\left (\mathbf{x}_{i} \right )$. There are several types of surrogate  techniques to choose from. Among the most popular ones are polynomial functions (e.g.\cite{queipo2005surrogate}), radial basis functions (e.g.\cite{wild2008orbit}), Kriging (e.g.\cite{kleijnen2009kriging}), support vector machines (SVMs) (e.g.\cite{stern2017accelerated,tabatabaee2012two}), and neural network (e.g.\cite{ziyadi2016efficient}).

\emph{Universal approximation theorem} in the mathematical theory of artificial neural networks rigorously proves that the standard multilayer, feed-forward, neural networks consist of one or more hidden layers with sufficiently many hidden units and, with arbitrary non-constant activation functions, can approximate any Borel-measurable function in a finite-dimensional space up to any arbitrary degree of accuracy \cite{hornik1989multilayer, hornik1991approximation}. This signifies that any failure in function approximation with sufficient accuracy by a multilayer network must be due to insufficient number of hidden units, inadequate learning, or lack of a deterministic input-output map \cite{hornik1989multilayer}. Although this theorem states that single-hidden-layer neural networks are already universal approximations, implementation of multiple layers will improve the performance of the neural network \cite{lecun2015deep}. With the cutting-edge neural network architectures and advanced training algorithms, \emph{deep learning} has recently been successfully used to solve elusive problems \cite{lecun2015deep} and have won several machine learning contests \cite{schmidhuber2015deep}. Deep learning consists of the development of computational models using  multiple processing layers in order to learn data representations with multiple abstraction levels \cite{lecun2015deep, goodfellow2016deep, demuth2014neural}. 

The goal of this paper is to propose a general framework  to accelerate reliability analysis of infrastructure systems. In this paper, we demonstrate how one can achieve this goal using deep neural network surrogates in the context of two-terminal reliability assessment of transportation networks subject to extreme earthquake events. Two distinct deep neural network surrogates are constructed and studied: a \emph{classifier} surrogate, which speeds up the two-terminal connectivity evaluation for a given network topology, and an \emph{end-to-end} surrogate that replaces the entire Monte Carlo simulation and can be used to immediately calculate the average (expected) two-terminal connectivity given the failure probability of network components. Although the idea of using artificial neural networks in reliability analysis of structures and infrastructure systems has been previously studied (e.g. \cite{cheng2008new,papadrakakis2002reliability,gomes2004comparison,hurtado2001neural,srivaree2002estimation,elhewy2006reliability,cardoso2008structural,zhang2004performance}), the major contributions of this work are as follows: (1) Neural network surrogates with multiple hidden layers were used to enhance the performance of surrogate-based two-terminal reliability analysis; (2) An end-to-end surrogate was proposed, which bypasses the sample-based calculations module that typically requires prohibitively large number of Monte Carlo simulations; and (3) In training the end-to-end surrogate, instead of using exact training data, we propose to use the predictions of the classifier surrogate to drastically reduce the computational time. We will numerically show that the proposed end-to-end surrogate is capable of accelerating the two-terminal reliability analysis of transportation networks by more than four orders of magnitude, and how such acceleration can substantially facilitate sensitivity analysis and potentially other planning procedures for large networks. 

The remainder of this paper is organized as follows. A general simulation-based framework for two-terminal reliability analysis of transportation networks subject to earthquake events is described in Section \ref{sec: reliability analysis}. Next, Section \ref{sec: deep learning} presents the proposed surrogate-based analysis of two-terminal reliability using deep neural networks. Finally, the accuracy and efficiency of the proposed surrogate-based analysis is demonstrated through a case study for the San Jose-Mountain view transportation network in Section \ref{sec: numerical examples}. 

\section{Two-Terminal Reliability Analysis} \label{sec: reliability analysis}

This section explains a general framework for two-terminal reliability analysis. First, the two-terminal connectivity of a network is introduced. Next, ground motion prediction equations are introduced, which enable the prediction of ground motion intensity measures at the location of network components. Given these predictions, it is then illustrated how one can evaluate the vulnerability of network components by the use of fragility analysis. Finally, a Monte Carlo simulation procedure is described for the analysis of system-level response.

\subsection{Two-Terminal Connectivity}

Consider a transportation network represented by a graph $\boldsymbol{G}=\left ( \boldsymbol{V},\boldsymbol{E} \right )$, where $\boldsymbol{V}$ is the set of nodes and $\boldsymbol{E}$ is the set of links (i.e. roadways). In the aftermath of an earthquake, the links connecting pairs of nodes $e_{ij}\in \boldsymbol{E}$, may stop functioning primarily due to bridge failures. The two-terminal connectivity is defined as follows. Given a source node $v_{s} \in \mathbf{V}$ and a terminal node $v_{t} \in \mathbf{V}$, the two-terminal connectivity is the condition where at least a connection exists between the source and terminal nodes. A pair of adjacent nodes $\left ( v_{i},v_{j} \right )$ are disconnected if there is at least one failed bridge on the link $e_{ij}$. In this work, it is assumed that bridges are the only components of the transportation network that are vulnerable to and get impacted by seismic hazards. This assumption is very common in the literature (e.g. \cite{bocchini2011stochastic}). The two-terminal connectivity problem is relevant when, for instance, the accessibility from a major attraction point to a major hospital, or from a feedstock to demand zones, is to be maintained during an emergency \cite{kang2008matrix}.

\subsection{Ground Motion Prediction}
For engineering applications, the evaluation of earthquake ground motions is generally performed using empirical Ground Motion Prediction Equations (GMPE) \cite{stewart2015selection, bommer2010selection}. GMPEs are statistical models that provide a means to predict the ground motion intensity measures, such as peak ground motions or response spectra, as a function of earthquake magnitude, source-to-site distance, fault mechanism, local site conditions, etc. GMPEs are generally constructed based on empirical data and are empirical regression models of recorded data. A summary of all the empirical GMPEs for estimation of earthquake Peak Ground Acceleration (PGA) and elastic response spectral ordinates published between 1964 and 2016 is provided in \cite{douglas2017ground}.

In this work, to determine the ground motion (specifically its spectral acceleration $S_{a}$) at a bridge site,  the Graizer-Kalkan 2015 (GK15) GMPE \cite{graizer2016summary, graizer2015update} is adopted. GK15 consists of predictive equations for spectral acceleration and PGA that are derived based on physical simulations and empirical data, which are applicable to earthquakes of moment magnitude $M$ between 5.0 and 8.0, at closest distances to fault rupture plane $R$ ranging from 0 to 200 km, at sites having $V_{s30}$ in the range of 200 to 1,300 m/s, and for spectral periods $T$ of 0.01$-$5 s. 
In GK15, the PGA, herein denoted by $ a_{\textup{PGA}}$, is calculated as a multiplication of a series of functions, and in natural logarithmic scale, is given by
\begin{equation} \label{PGA}
\ln\left (a_{\textup{PGA}}  \right ) =\ln\left (G_{1}  \right )+\ln\left (G_{2}  \right )+\ln\left (G_{3}  \right )+\ln\left (G_{4}  \right )+\ln\left (G_{5}  \right )+\sigma _{\ln\left ( a_{\textup{PGA}} \right )},
\end{equation}
where $G_{1}$ represents a scaling function for magnitude and style faulting, $G_{2}$ is a model for ground motion attenuation, $G_{3}$ is a model for adjustment to the attenuation rate in order to take into account the regional anelastic attenuation, $G_{4}$ represents the site amplification model, and $G_{5}$ represents a model for basin scaling. $\sigma _{\ln\left ( a_{\textup{PGA}} \right )}$ is the residual variability, which accounts for unexplained variability in the ground motion data used for the calibration of GMPE. In seismic hazard analysis, reducing this residual variability is of a high priority, since at large values of $a_{\textup{PGA}}$, the probabilities of exceedance go up rapidly with $\sigma _{\ln\left ( a_{\textup{PGA}} \right )}$. As will be shown in the numerical examples, the two-terminal connectivity of  San Jose-Mountain View transportation network is significantly affected by this residual variability \cite{graizer2016summary}. 

The form of GK15 for the 5\% damped $S_{a}$ response ordinates is
\begin{equation} \label{Sa}
S_{a}=a_{\textup{PGA}} \ \  \mu \left ( M,R,V_{\textup{s30}},B_{\textup{depth}} \right ),
\end{equation}
where the spectral shape $\mu$ is parameterized by $M$, $R$, $V_{s30}$, and basin depth under the site $B_{\textup{depth}}$. For the analysis of bridge fragility, as described in the next subsection, spectral accelerations at 0.3~s and 1.0~s are used.

\subsection{Bridge Fragility Analysis}

There are several well-established ways for the analysis of structural response to natural hazards. In this study, the HAZUS-MH fragility model \cite{fema2008hazus}, developed by the Federal Emergency Management Agency (FEMA), is implemented for the calculation of transportation network bridge response to earthquake ground shaking. HAZUS-MH is a standardized methodology for the estimation of potential physical, economic, and social losses from earthquakes, hurricanes, and floods. For a given level of ground motion, fragility curves or damage functions for bridges are modeled as functions with log-normal distributions that yield the probability of reaching or exceeding different damage states. Individual fragility curves are parametrized by a median value of ground motion or ground failure, and an associated standard deviation.

The required inputs needed to estimate the damages to a bridge in HAZUS-MH fragility model are geographical location of the bridge (latitude and longitude), spectral accelerations at $0.3$ s and $1.0$ s at the bridge location, peak ground acceleration, soil type, and bridge classification. Bridges are classified into 28 primary types based on several structural characteristics, such as seismic design, structure type, number of spans, and pier type. Five damage states are considered for bridges, which are none, slight, moderate, extensive, and complete damage states. Extensive damage for bridges is defined by shear failure, degradation of columns with no collapse, differential settlement at connections, large residual movement at connections, and shear key failure at abutments.  In this study, it is assumed that the bridges will stop functioning at the onset of extensive damage state immediately after an earthquake event.

For each of the bridge classes, a total of four different fragility curves are constructed from the combination of two log-normal distributions for ground shaking and ground failure. Afterward, specific fragility curves for individual bridges are constructed by updating the generic curves based on the bridge characteristics. The output of fragility analysis for each bridge is four different curves that represent the probability of that bridge exceeding a damage state for a given level of ground motion. These fragility curves are then used, as illustrated in the next subsection, in order to calculate the system-level response of transportation network to an earthquake via a simulation-based study of two-terminal connectivity.

\subsection{Two-Terminal Reliability Analysis}
 In order to estimate the system-level network response to an earthquake affecting it components, e.g. roadways, Monte Carlo Simulation (MCS) may be used. MCS is a straight-forward, easy to implement, approach ideally suited to parallel computing. For calculation of the  two-terminal connectivity in this study, network realizations are drawn by randomly removing roadways  according to their survival probabilities, given by Equation~\ref{link_fail_prob}.  Specifically, the damage state for each roadway is modeled as a Bernoulli random variable with the following distribution
\begin{equation} \label{Bernoulli}
x_{i} = \left\{\begin{matrix}
1, & \! \! \! \! \! \!  \textup{with probability} \,\, \, p_{i}, \\ 
0, & \; \; \; \textup{with probability} \,\, \, 1-p_{i},
\end{matrix}\right.
\end{equation}
where $\{ 0,1  \}$ denotes the survived and failed states, respectively, with a survival probability of $p_{i}$.
A roadway with at least one failed bridge will be removed. Therefore, the survival probability $p_{i}$ of roadway $i$ with $k$ bridges of IDs $\left \{ i_1,\dots,i_k  \right \}$ is calculated in logarithmic scale as
\begin{equation} \label{link_fail_prob}
\ln\left ( p_i \right ) = \sum_{j=1}^{k}\left [ \ln\left (  {p}_{i_j} \right ) \right ],
\end{equation}
where $ {p}_{i_j}$ is the survival probability of bridge $i_j$.

Let $v_{s} \in \mathbf{V}$ and $v_{t} \in \mathbf{V}$ denote, respectively, the source and terminal nodes predetermined by the stakeholder. For a network realization using the $j^{th}$ MC sample, the two-terminal connectivity is assessed by evaluating whether there is any connection between the source and terminal
\begin{equation} \label{Connectivity}
g_{j}\left ( x_{1}^{(j)},x_{2}^{(j)},...,x_{\ell}^{(j)} \right )=\left\{\begin{matrix}
1, &\; \; \;\;   \textup{if s,t are connected}, \\ 
0, &\! \! \! \! \! \!\! \!\! \! \! \! \! \! \! \!\!     \textup{otherwise},
\end{matrix}\right.
\end{equation}
in which $\ell$ is the total number of roadways in the network. This procedure is repeated by drawing more network realizations until convergence of the quantity of interest (QoI) is achieved. The Depth-First Search (DFS) algorithm \cite{tarjan1972depth,korf1985depth} with a linear-time computational complexity of $\bigO \left ( \left | V \right |+\left | E \right | \right )$ is utilized herein for the evaluation of two-terminal connectivity. For a given set of failure probabilities for bridges, and for a given MCS with $N$ network realizations, the  expected two-terminal connectivity $P_c$ is estimated by
\begin{equation} \label{MCS}
P_c=\frac{1}{N}\sum_{j=1}^{N}g_{j}(x_{1}^{(j)},x_{2}^{(j)},...,x_{\ell}^{(j)}),
\end{equation}

In order to accelerate the two-terminal connectivity computations, the Monte Carlo calculations are performed in parallel  \cite{rosenthal2000parallel} where different processors evaluate the network connectivity for different network realizations. In the next section, we explain the approach to train and use fast and accurate deep learning surrogates in place of Monte Carlo-based DFS (exact) calculations.

\section{Surrogate Model for Two-Terminal Connectivity} \label{sec: deep learning}

\subsection{Deep Neural Networks}
For notation brevity, \textit{single hidden layer} neural networks are introduced first, since its subsequent generalization to multiple hidden layers, which makes a neural network \emph{deep}, will be straightforward. Given the $d$-dimensional row vector $\mathbf{x} \in D^{d}$ as model input, the $k$-dimensional output of a standard single hidden layer neural network is in the form of
\begin{equation} \label{OHL-NN}
\textbf{y} = \sigma \left (\textbf{x} \textbf{W}_{1}+\textbf{b}_{1}  \right ) \textbf{W}_{2}+\textbf{b}_{2},
\end{equation}
in which ${W}_{1}$ and ${W}_{2}$ are weight matrices of size $d\times q$ and $q\times k$, respectively, ${b}_{1}$ and ${b}_{2}$ are biases of size $1\times q$ and $1\times k$, respectively. The function $\sigma\left ( \cdot \right )$ is an element-wise non-linearity, commonly known as the \textit{activation} function. In deep neural networks, the output of each activation function is transformed by a new weight matrix and a new bias, and is then fed to another activation function. For each new set of weight matrix and bias that is added to (\ref{OHL-NN}), a new \textit{hidden layer} is added to the neural network. The capacity of neural networks can be easily increased by adding more hidden layers or more units to each hidden layer.

Popular choices of activation functions are Sigmoid, hyperbolic tangent (Tanh), and rectified linear unit (RELU). The RELU activation function has the form of $f\left ( \theta  \right )=\max\left ( 0,\theta  \right )$. RELUs are getting increasingly popular in deep learning applications as, compared to Sigmoid and Tanh activations, they are faster and do not suffer from the vanishing gradient problem.

In order to calibrate the weight matrices and biases for a regression problem, we may use a Euclidean loss function as follows
\begin{equation} \label{MSE Loss}
E_\textup{MSE}\left ( \textbf{X},\textbf{Y}\right)=\frac{1}{2M}\sum_{i=1}^{M}\left \| \textbf{y}_i-\hat{\textbf{y}}_{i} \right \|^{2},
\end{equation}
where $E_\textup{MSE}$ is the mean squared error, $X=\left \{ \textbf{x}_1,\textbf{x}_2,...,\textbf{x}_M \right \}$ is the set of $M$ observed inputs, $Y=\left \{ \textbf{y}_1,\textbf{y}_2,...,\textbf{y}_M \right \}$ is the set of $M$ observed outputs, and $\left \{ \hat{\textbf{y}}_1,\hat{\textbf{y}}_2,...,\hat{\textbf{y}}_M \right \}$ is the set of neural network output (model prediction) corresponding to the set of inputs $X$. For a binary classification task, we may use a binary cross-entropy loss function in the form of
\begin{equation} \label{Binary Cross-Entropy Loss}
E_\textup{BCE}\left ( \textbf{X},\textbf{Y}\right)=-\sum_{i=1}^{M} \{ \textbf{y}_{i}\log\left ( \hat{\textbf{y}}_{i} \right )+\left ( 1-\textbf{y}_{i} \right )\log\left (  1-\hat{\textbf{y}}_{i}  \right ) \},
\end{equation}
where $E_\textup{BCE}$ is the binary cross-entropy. Minimizing the loss function with respect to model parameters ($\mathbf{W}_{1},\mathbf{W}_{2},\cdots,\mathbf{b}_{1},\mathbf{b}_{2}$) will yield the calibrated model parameters ($\mathbf{W}_{1}^{*},\mathbf{W}_{2}^{*},\cdots,\mathbf{b}_{1}^{*},\mathbf{b}_{2}^{*}$). For instance, for a binary cross-entropy loss function, we have

\begin{equation} \label{minimize_loss}
\left ( \mathbf{W}_{1}^{*},\mathbf{W}_{2}^{*},\cdots,\mathbf{b}_{1}^{*},\mathbf{b}_{2}^{*},\cdots  \right )=\underset{{\left ( \mathbf{W}_{1},\cdots,\mathbf{b}_{1}\cdots  \right )}}{\operatorname{argmax}} E_\textup{BCE}\left (\textbf{X},\textbf{Y}\right),
\end{equation}

Minimizing the loss function is usually performed using \emph{backpropagation} \cite{lecun2015deep,jin2000improvements}. It consists of a two-phase cycle; \emph{forward pass} and \emph{backward pass}. A forward pass takes the input to the network and propagates it through the layers of the network, one by one, to calculate the network output. A backward pass starts from the network output and propagates towards the input layer while calculating the gradients, layer by layer, using the chain rule.

\subsection{Deep Neural Networks for Two-Terminal Reliability Analysis}\label{subsec:DNN_reliability}
The step-by-step procedure for construction of DNN surrogates that can be used to accelerate two-terminal reliability analysis of transportation systems is elaborated in this section. Two different surrogate models are developed in this study. The first model is hereinafter  referred to as the \emph{classifier surrogate}. It replaces the DFS algorithm  to determine whether a particular source-to-terminal connection exists. It does so for each MC sample, i.e. for each realized roadway failure and its corresponding topology. The input to this model is therefore a deterministic network topology in the form of a binary vector, and the output is a binary variable indicating the connection. 

\begin{figure}
	\centering
	\includegraphics[width=0.95\linewidth]{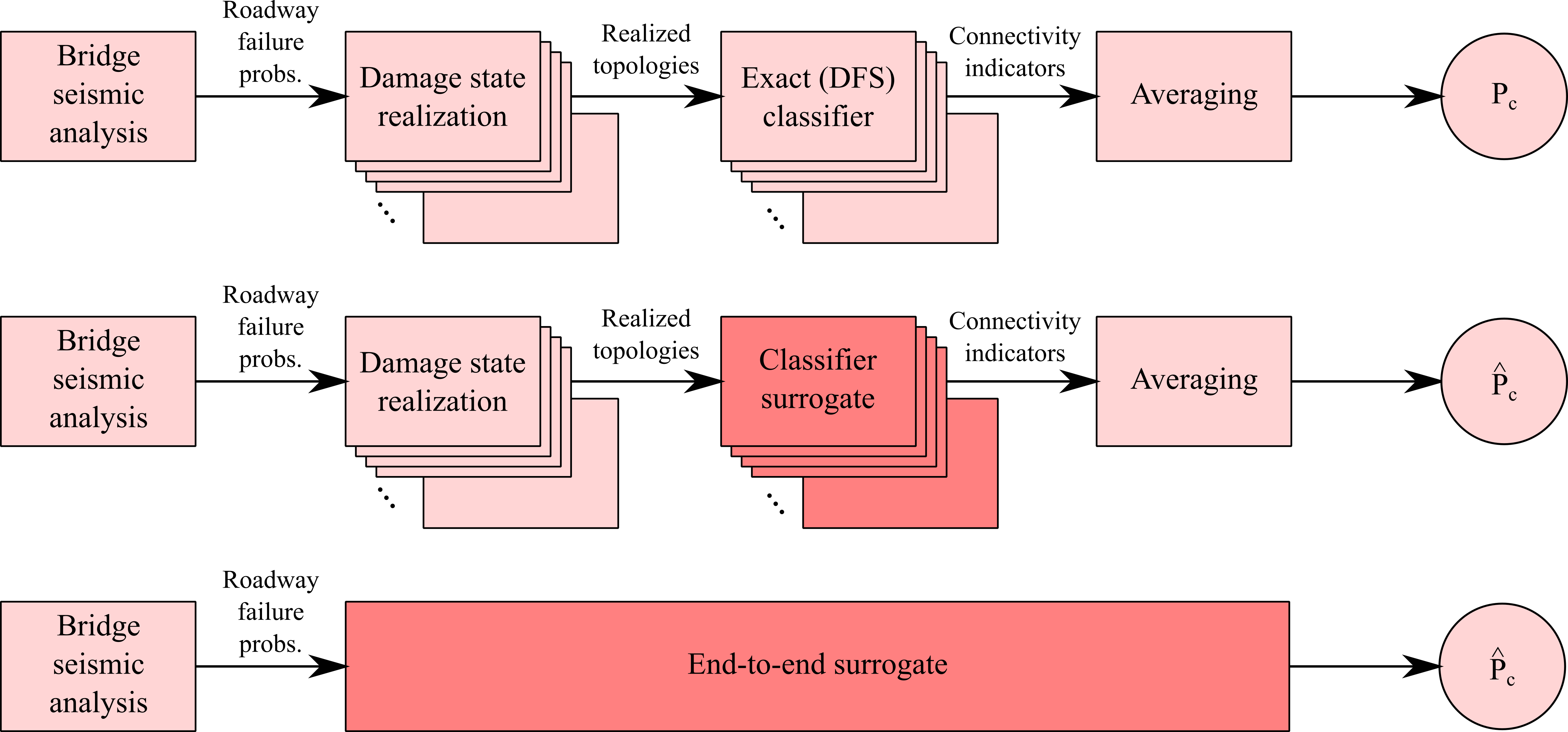}
	\captionsetup{}
	\caption{Workflow for calculation of the expected two-terminal connectivity, using exact (DFS) connectivity check (top), classifier surrogate (middle), and end-to-end surrogate (bottom).} 
	\label{workflow}
\end{figure}

\begin{figure}
	\centering
	\includegraphics[width=1.0\linewidth]{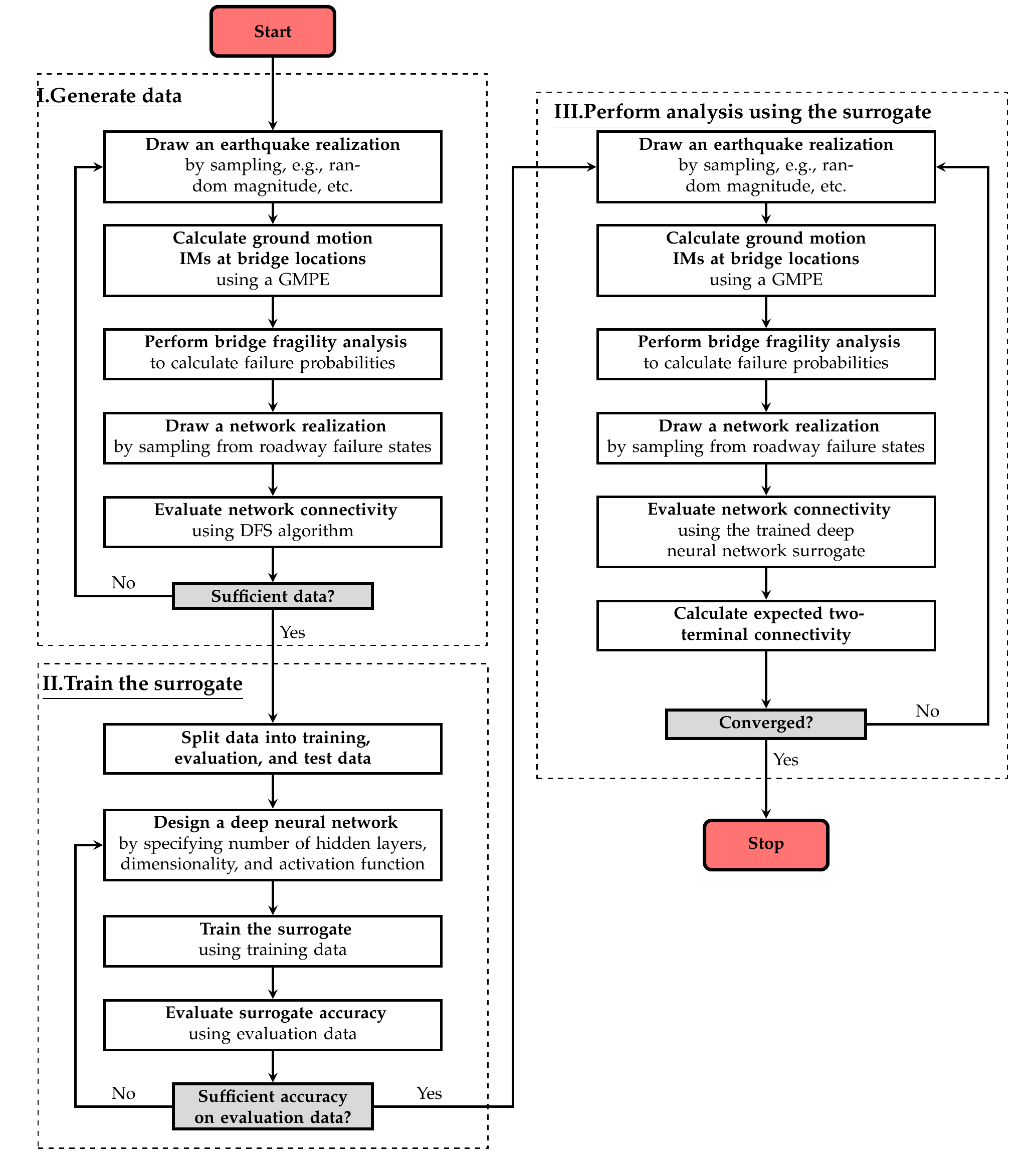}
	\captionsetup{}
	\caption{Framework for constructing the classifier surrogate and utilizing it for Monte Carlo-based two-terminal reliability analysis. In this procedure, the classifier surrogate will replace the DFS algorithm.}
	\label{flowchart1}
\end{figure}

\begin{figure}
	\centering
	\includegraphics[width=0.45\linewidth]{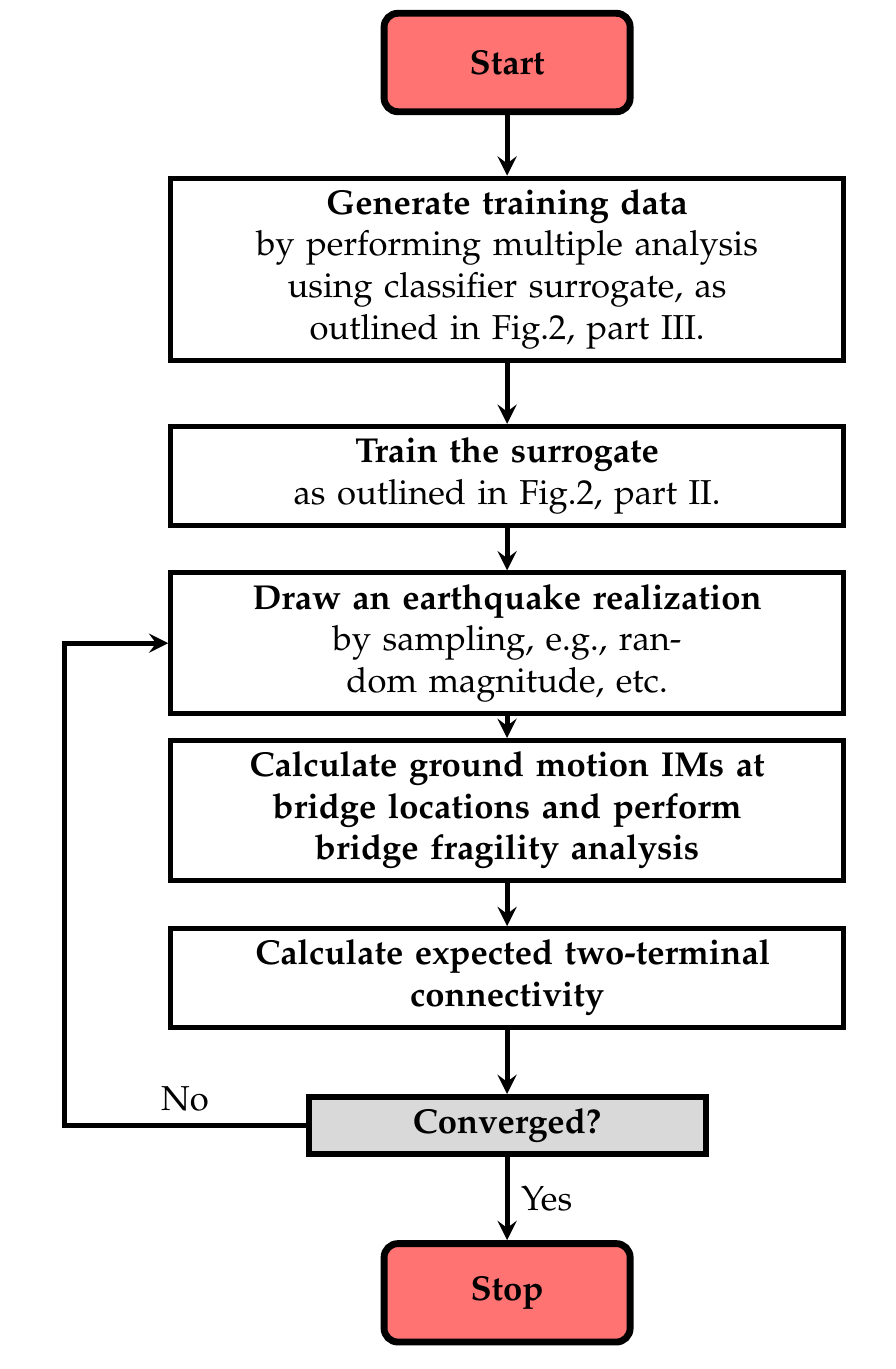}
	\captionsetup{}
	\caption{Framework for constructing the end-to-end surrogate and utilizing it for two-terminal reliability analysis with end-to-end surrogate replacing  Monte Carlo simulations.} 
	\label{flowchart2}
\end{figure}

The second surrogate model, which we refer to as the \emph{end-to-end surrogate}, is designed to replace the topology realization, connectivity determination, and connectivity averaging modules (see Figure~\ref{workflow}). It is used to immediately evaluate the average (expected) two-terminal connectivity given the roadway failure probabilities. It  bypasses  roadway status realizations from the failure  probabilities, and thus saves computational time. Figures \ref{flowchart1}  and \ref{flowchart2} show the proposed frameworks for constructing the classifier and end-to-end surrogates, respectively,  and  how these surrogates are utilized in the evaluation of expected two-terminal connectivity. 

\subsection{Surrogate Performance Measures}

In order to evaluate the accuracy of a DNN surrogate model, a number of performance measures are used in this study. They include QoI prediction accuracy $\alpha_\textup{QoI}$, binary classification accuracy $\alpha_\textup{binary}$, sensitivity or True Positive Rate (TPR), and specificity or True Negative Rate (TNR) \cite{baldi2000assessing}. The last three measures are applicable to binary classification only.

The QoI prediction accuracy for  connectivity is calculated as
\begin{equation} \label{QoI accuracy}
\alpha_\textup{QoI}=1-\frac{\left | P_c-\hat{P}_c \right |}{P_c},
\end{equation}
where $P_c$  and $\hat{P}_c$ are the two-terminal connectivity calculated respectively using exact (DFS) connectivity check and the surrogate. The binary classification accuracy is calculated as
\begin{equation} \label{binary accuracy}
\alpha_\textup{binary}=\frac{TP+TN}{TP+FP+TN+FN},
\end{equation}
where TP (True Positive) is the number of times the surrogate correctly predicts network survival, TN (True Negative) is the number of times the surrogate correctly predicts network failure, FP (False Positive) is the number of times the surrogate incorrectly predicts network survival, and FN (False Negative) is the number of times the surrogate incorrectly predicts network failure, satisfying $TP+TN+FP+FN=N$. As more specific measures, True Positive Rate (TPR) and True Negative Rate (TNR) are calculated as

\begin{eqnarray}
TPR=\frac{TP}{TP+FN},     \label{sensitivity} 
\\ \nonumber \\
TNR=\frac{TN}{TN+FP}.   \label{specificity}
\end{eqnarray}

\section{Case Study for the San Jose-Mountain View Transportation Network} \label{sec: numerical examples}

\begin{figure}
	\centering
	\includegraphics[width=0.99\linewidth]{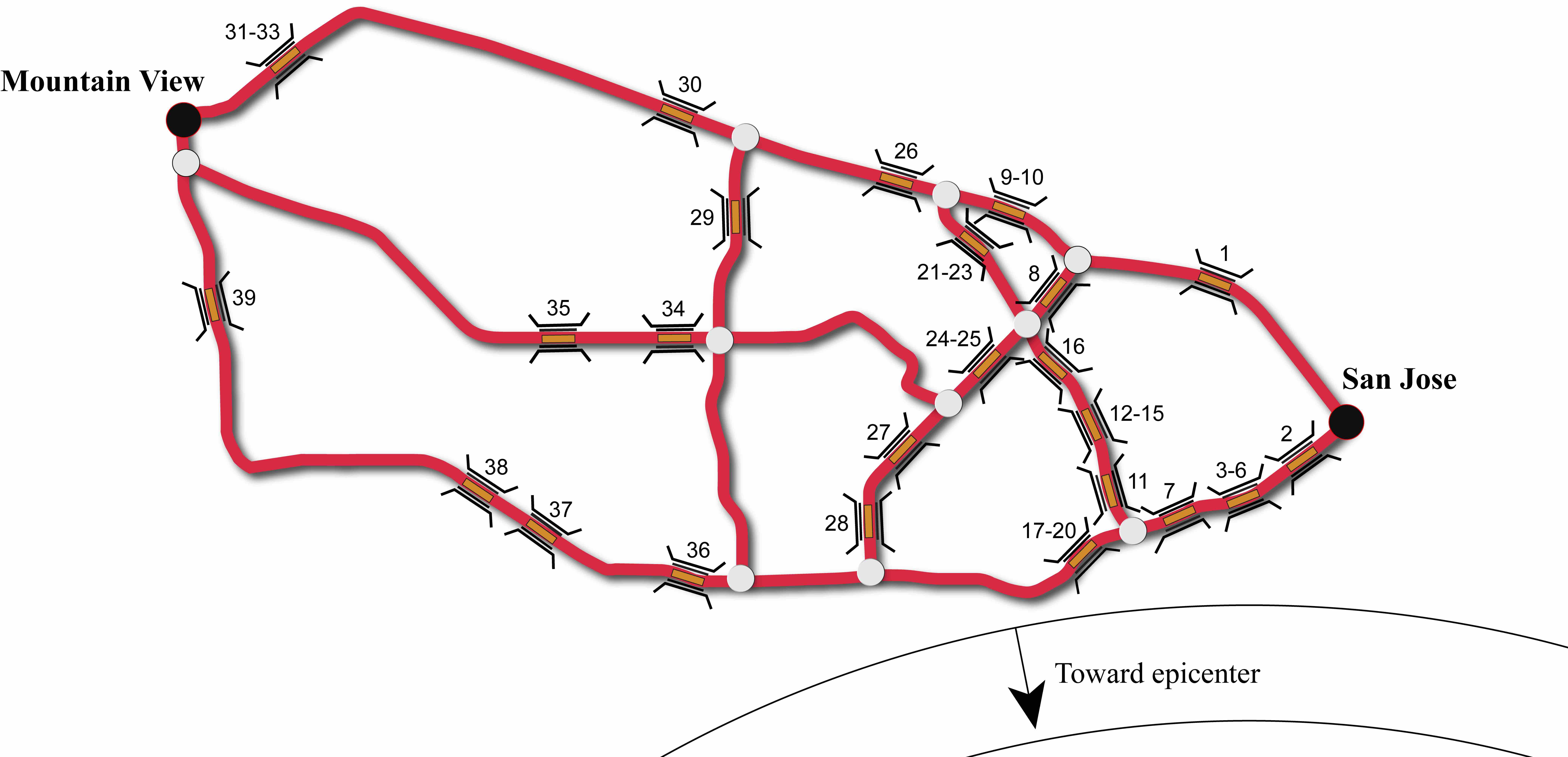}
	\captionsetup{}
	\caption{Layout of the San Jose-Mountain View transportation network. Numbers near each bridge are the ID of that bridge. Multiple IDs shows that there are multiple bridges there but only one is shown.} 
	\label{fig.network}
\end{figure}

The surrogate-based two-terminal connectivity analysis procedure described in Section \ref{subsec:DNN_reliability}   is applied to the transportation network that connects San Jose, CA to Mountain View, CA in the United States. This network is located in a region of high seismic activity. A sketch of this network is provided in Figure \ref{fig.network}. The network  consists of 39 bridges, 12 nodes, and 18 roadway links (out of which 14 have at least one bridge). Throughout the numerical examples presented in this section,  the network is considered to be impacted by the 1989 Loma Prieta earthquake with varying magnitudes. The geographical coordinates for the epicenter of Loma Prieta earthquake are $37.04^\circ\textup{N}$, $121.88^\circ\textup{W}$.

A number of assumptions and choices were made throughout this section. First,  bridges are assumed to be the only network components  vulnerable to earthquakes, as commonly considered in the literature (e.g. \cite{bocchini2011stochastic}). Second, the network is considered to be an undirected graph since the adjacent bridges on two different sides of the road share the same or very similar properties. Also, according to the HAZUS-MH soil classification, the soil for the study area is determined to be of type D. 

The NetworkX Python library \cite{schult2008exploring}  was used for network connectivity evaluation using DFS algorithm,  and the Keras deep learning library \cite{chollet2015keras} was used for construction of classifier and end-to-end surrogates. The Python source codes for all the simulations presented in this section are made available on GitHub \cite{mohammad_amin_nabian_2017_846898}. Computations in sections \ref{example1}, \ref{example2} are conducted on a quad core 2.5 GHz MacBook Pro with 16 GB of RAM. Computations in sections \ref{example3}, \ref{example4} are conducted on the Comet cluster from XSEDE (Extreme Science and Engineering Discovery Environment) resources [50], with 24 2.5 GHz CPU cores and 128 GB of DRAM.

\subsection{Classifier Surrogate Training and Prediction} \label{example1}
\begin{figure}
	\centering
	\includegraphics[width=0.5\linewidth]{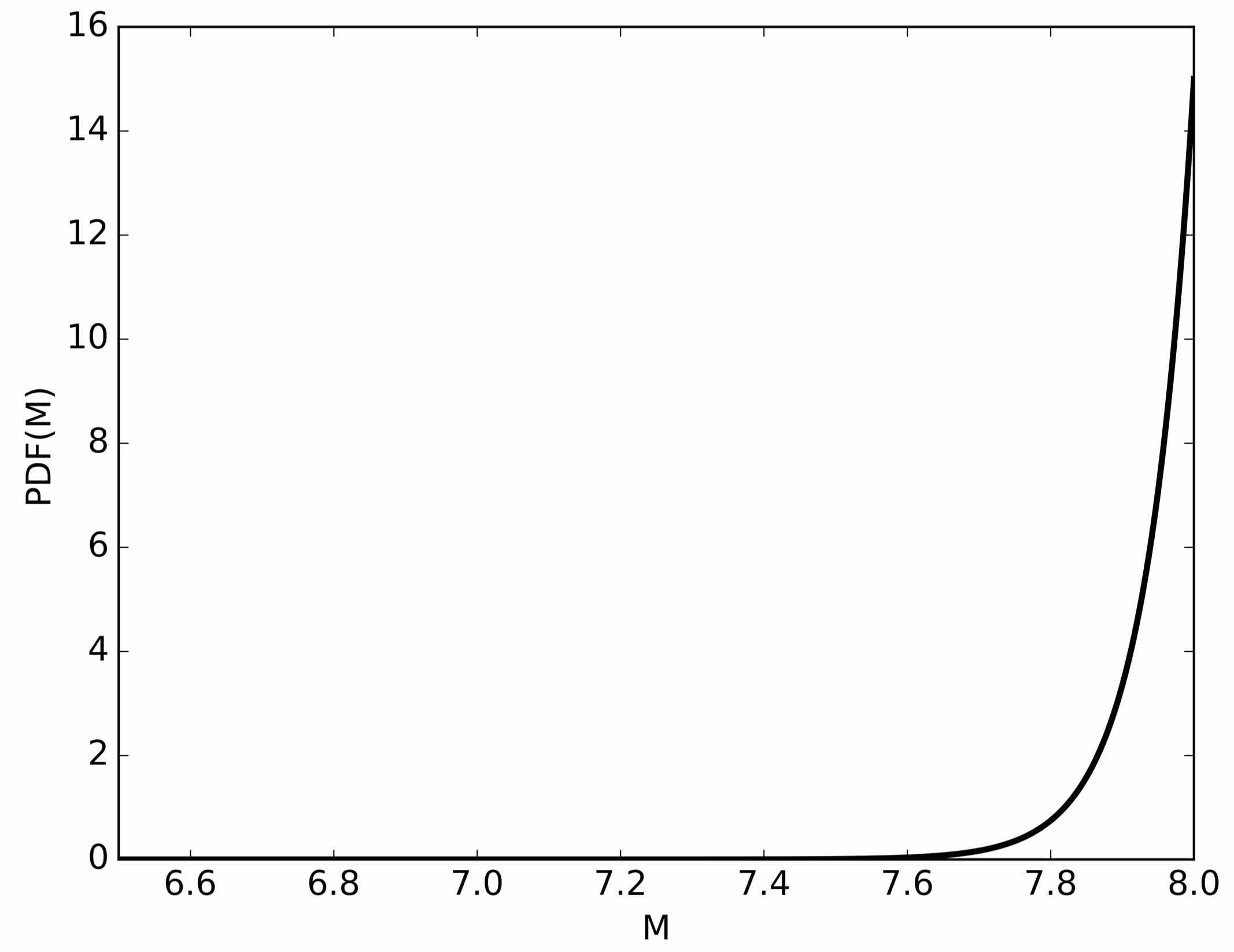}
	\captionsetup{}
	\caption{The probability density function used to draw samples from earthquake magnitude in order to generate surrogate training and evaluation data sets.} 
	\label{Magnitude_plot}
\end{figure}

Following the framework represented in Figure \ref{flowchart1}, given a network realization, the classifier surrogate indicates whether a source-to-terminal connection exists. The input and output to this surrogate are a binary vector of roadway conditions (failed or survived) and a binary connectivity indicator, respectively. In order to generate training and evaluation data sets, a total of 10,000 samples of earthquake magnitude $M$, denoted by $\{m_i \}_{i=1}^{10000} $ are drawn according to 
\begin{equation} \label{example1_magnitude}
m_{i} = 8.0 - \theta_{i},
\end{equation}
where $\theta_{i}$ is a random sample drawn from a truncated exponential distribution with a shape parameter and lower and upper bounds of 15, 0, and 1.5, receptively. Ninety percent of the samples are used for training, and the rest is left for surrogate evaluation. A sketch for the probability distribution of $M$ is provided in Figure \ref{Magnitude_plot}. Training and evaluation samples are preferred to be drawn from an exponential distribution, and not a uniform distribution. This is due  to the non-linear relationship between earthquake moment magnitude and energy release \cite{kanamori1977energy}, leading to larger sensitivity  of failure probabilities to magnitude perturbations when the nominal magnitude is larger; hence, the exponentially increasing distribution of training samples.

\begin{figure}
 	\centering
 	\includegraphics[width=1.0\linewidth]{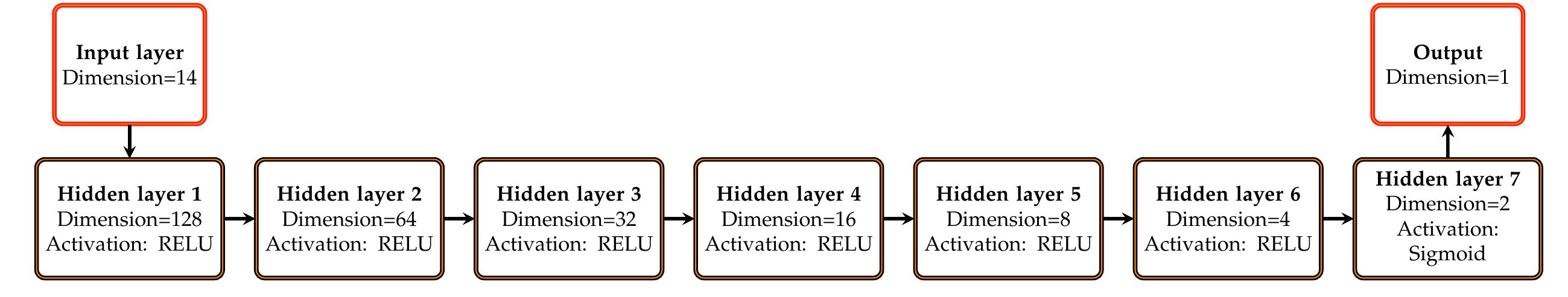}
	\captionsetup{}
	\caption{Architecture of the classifier surrogate. This surrogate is to be used instead of the DFS algorithm in order to accelerate network connectivity evaluation given a network realization. The model consists of 7 hidden layers with different dimensionalities. The input and output to this model are, respectively, a binary vector of roadway failure states and a scalar that represents the expected two-terminal connectivity.} 
	\label{Architecture1}
\end{figure}

The classifier surrogate consists of 7 hidden layers with different dimensionalities (see Figure \ref{Architecture1}). RELU activation is adopted for hidden layers 1 through 6, while the Sigmoid activation is used in the last hidden layer. The Adam optimization algorithm \cite{kingma2014adam}is used to minimize the binary cross-entropy loss function (Equation \ref{Binary Cross-Entropy Loss}). For 150 epochs and a batch size of 64, it took 83.05 seconds to train the classifier surrogate.

\begin{figure}
	\centering
	\includegraphics[width=0.6\linewidth]{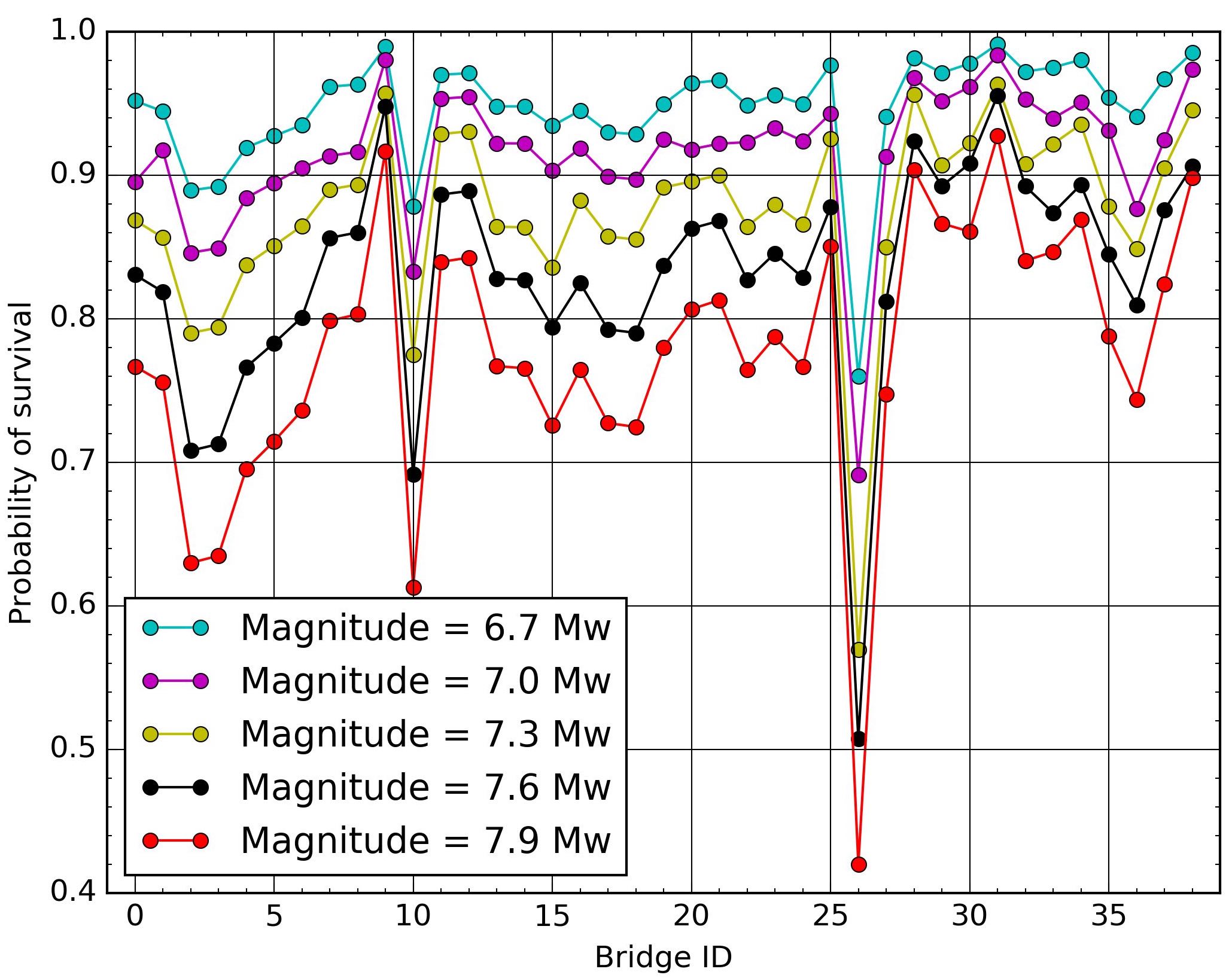}
	\captionsetup{}
	\caption{Bridge survival probabilities for different earthquake events considered in example 1.} 
	\label{fig.bridge_survavial_probs}
\end{figure}

\begin{figure}
	
	\begin{subfigure}[t]{0.499 \linewidth}
		\includegraphics[width=.99\linewidth]{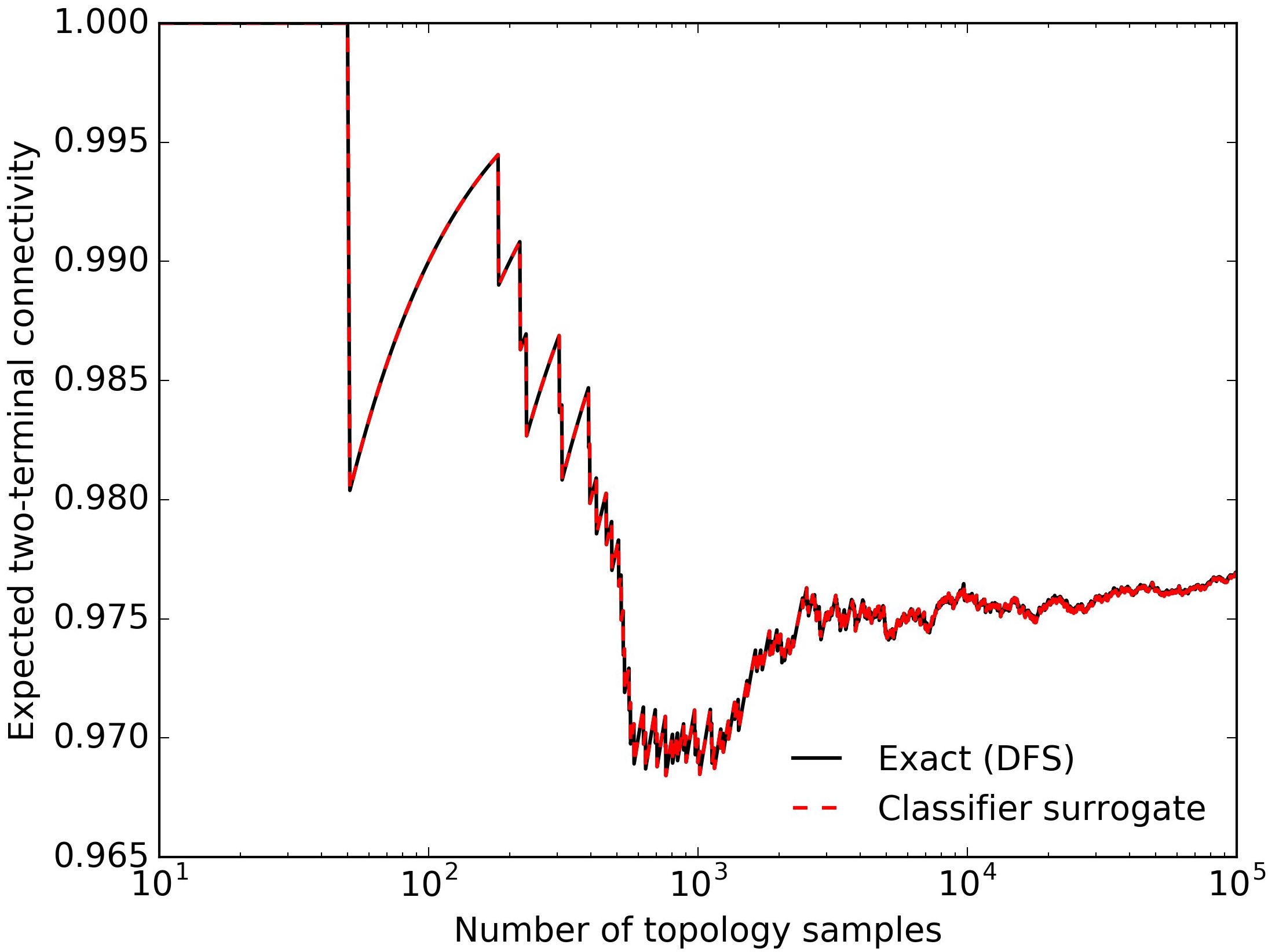}
		\caption{}
		\label{fig:Convergence_Plot_6.7}	
		\centering	
	\end{subfigure}
	\quad
	\begin{subfigure}[t]{0.499 \linewidth}
		\includegraphics[width=.99\linewidth]{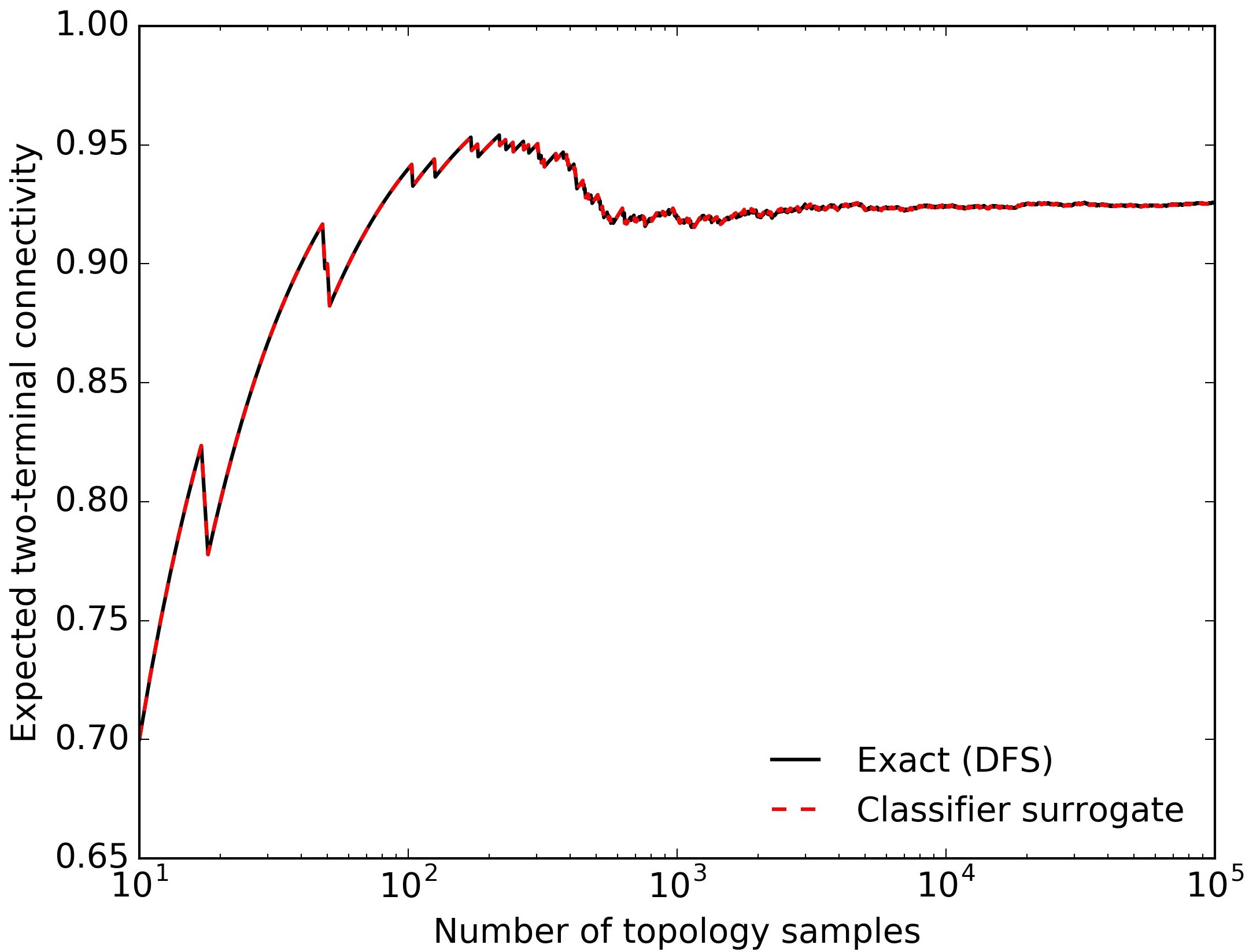}
		\caption{}
		\label{fig:Convergence_Plot_7.0}
	\end{subfigure}
	\\
	\begin{subfigure}[t]{0.499 \linewidth}
		\includegraphics[width=.99\linewidth]{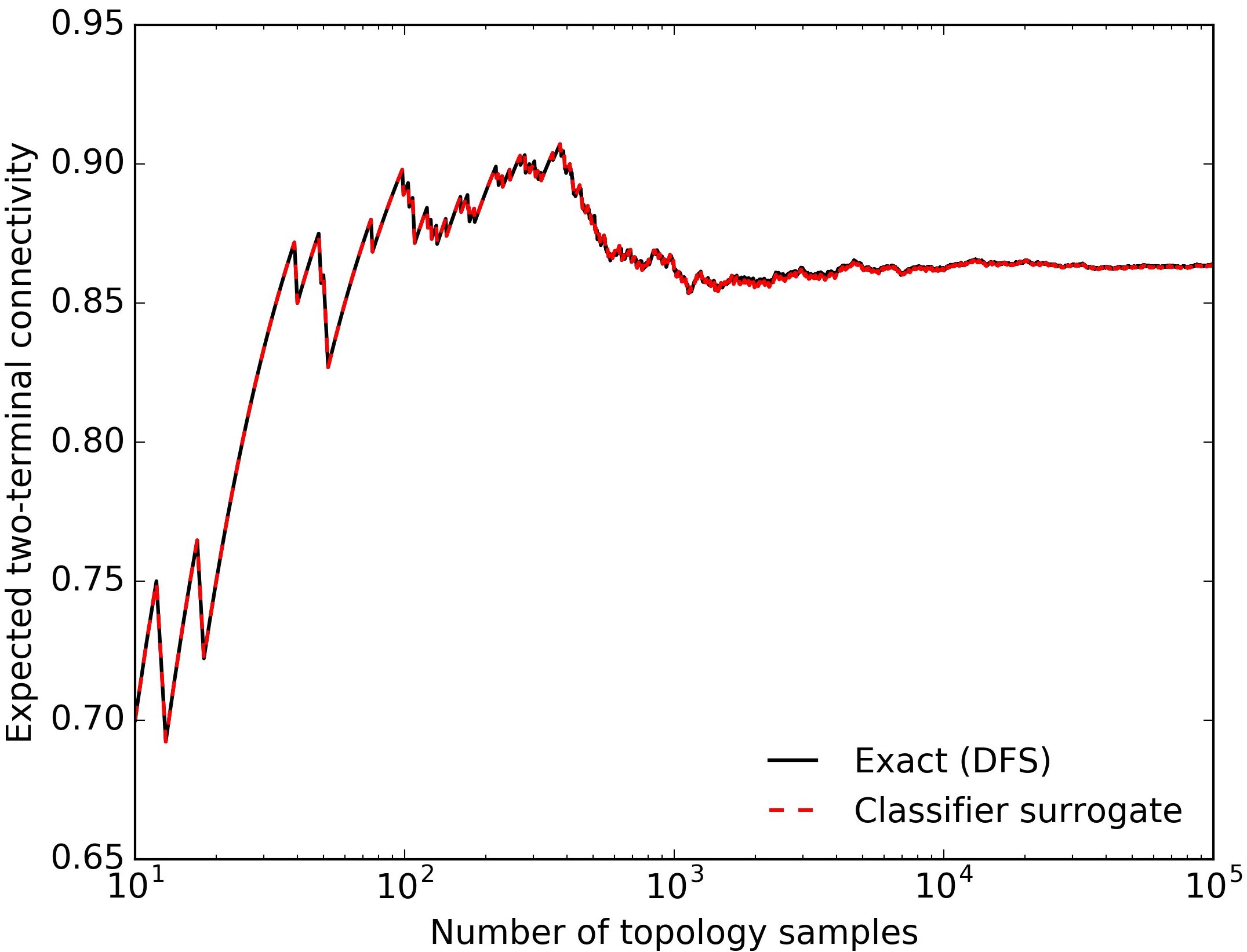}
		\caption{}	
		\label{fig:Convergence_Plot_7.3}
	\end{subfigure}
	\quad
	\begin{subfigure}[t]{0.499 \linewidth}
		\includegraphics[width=.99\linewidth]{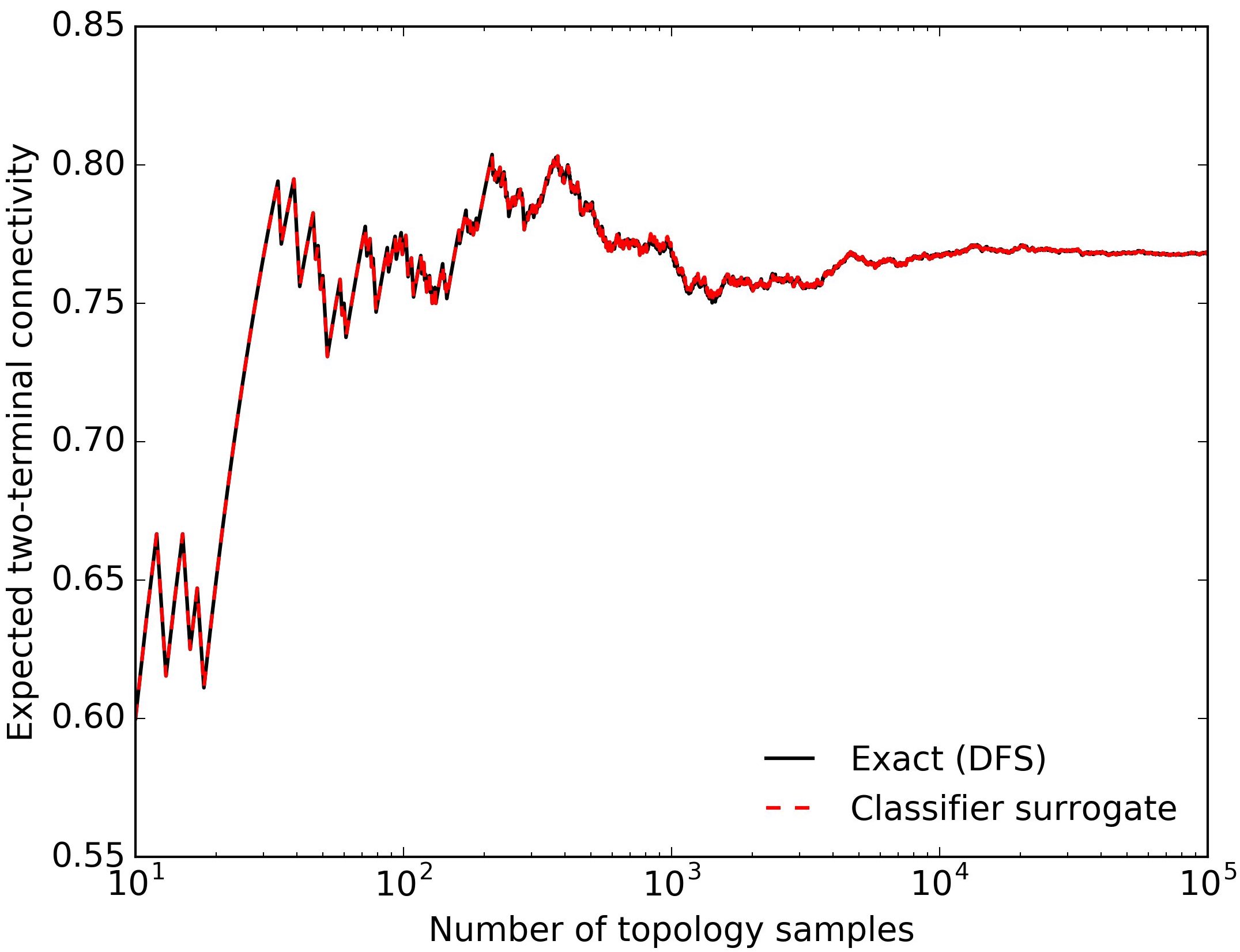}
		\caption{} 
		\label{fig:Convergence_Plot_7.6}
	\end{subfigure}
	\
\begin{subfigure}[t]{\linewidth}
	\centering
	\includegraphics[width=.499\linewidth]{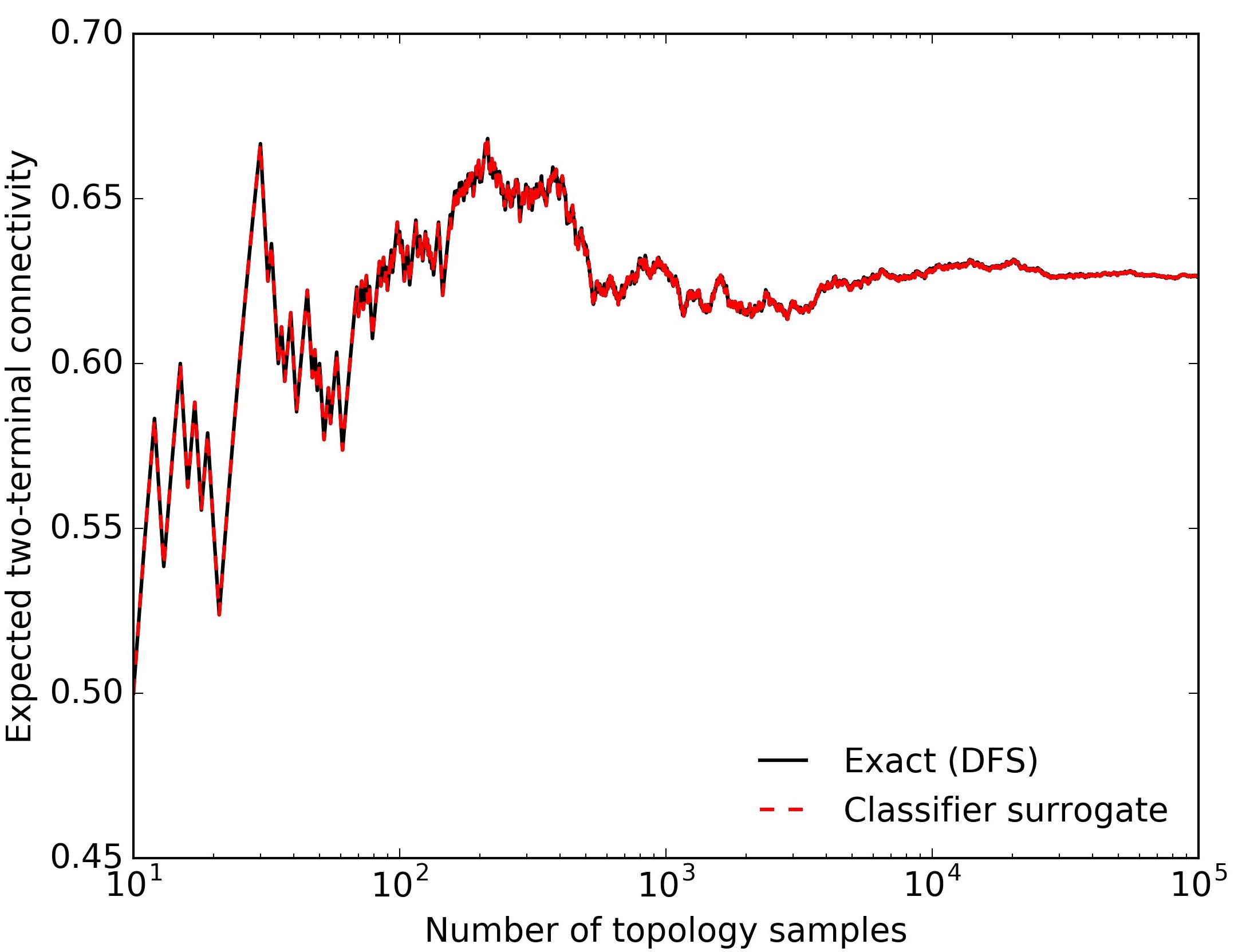}
	\caption{}	
	\label{fig:Convergence_Plot_7.9}
\end{subfigure}
	\captionsetup{}
	\caption{Convergence plots for prediction of two-terminal connectivity for different earthquakes of magnitude (a) 6.7 Mw, (b) 7.0 Mw, (c) 7.3 Mw, (d) 7.6 Mw, (e) 7.9 Mw. The x-axis is in logarithmic scale.} 
	\label{fig:example1_convergence_plot}
\end{figure}

In order to evaluate the predictive performance of the trained classifier surrogate, we consider five different scenarios with earthquake  magnitudes 6.7, 7.0, 7.3, 7.6, and 7.9 $\textup{M}_\textup{w}$, and for each we use the trained surrogate for two-terminal connectivity evaluation. The surrogate-based results are compared versus exact connectivity results obtained using DFS algorithm. Figure \ref{fig.bridge_survavial_probs} shows the  survival probabilities for the 39 bridges subject to the five earthquake scenarios. For each earthquake event, given these survival probabilities and by using Equation \ref{link_fail_prob} for calculating roadway failure probabilities, a total of 100,000 network realizations are generated. The two-terminal connectivity of each one of these network realizations is determined using the classifier surrogate and the DFS algorithm, and the resulting expected connectivities are compared in Figure \ref{fig:example1_convergence_plot}. It is evident from the convergence plots that the surrogate and DFS results are in close agreement. The estimated expected values for two-terminal connectivity, as well as computational times, are compared  in Table \ref{table:performance}, and surrogate performance measures are  reported in Table \ref{table:accuracy}. Compared to exact connectivity check using DFS, the classifier surrogate predictions are about one order of magnitude faster, with accuracies of more than 99.9\%.

\begin{table}[H]
	\centering
	\captionsetup{}
	\caption{ Two-terminal connectivity predictions and computational times for different earthquake magnitudes using DFS and classifier surrogate.}
	\label{table:performance}
	\scalebox{0.9}{
		\begin{tabular}{|M{2.5cm}|M{2.5cm}|M{2.5cm}|M{2.5cm}|M{2.5cm}|}
			\hline
			\textbf{Magnitude} &
			\textbf{$P_c$} & \textbf{$\hat{P}_c$}  &\textbf{DFS time (s)}  & \textbf{Surrogate time (s)}   \\ \thickhline
			\textbf{\begin{tabular}[c]{@{}c@{}} 6.7 \end{tabular}} & {0.9769} & 0.9769  & 5.23 & 0.62 \\ \hline
			\textbf{\begin{tabular}[c]{@{}c@{}} 7.0 \end{tabular}} & {0.9256} & 0.9255  & 5.40 & 0.62 \\ \hline
			\textbf{\begin{tabular}[c]{@{}c@{}} 7.3 \end{tabular}} & {0.8635} & 0.8633  & 6.05 & 0.63 \\ \hline
			\textbf{\begin{tabular}[c]{@{}c@{}} 7.6 \end{tabular}} & {0.7679} & 0.7679  & 5.71  & 0.62 \\ \hline
			\textbf{\begin{tabular}[c]{@{}c@{}} 7.9 \end{tabular}} & {0.6263} & 0.6264  & 5.63 & 0.65 \\ \hline
	\end{tabular}}
	
\end{table}

\begin{table}[H]
	\centering
	\captionsetup{}
	\caption{Classifier surrogate accuracy indicators for different earthquake magnitudes}
	\label{table:accuracy}
	\scalebox{0.9}{
			\begin{tabular}{|M{2.5cm}|M{2.5cm}|M{2.5cm}|M{2.5cm}|}
			\hline
			\textbf{Magnitude} & \textbf{$\mathbf{\alpha_{binary}}$} & \textbf{ TPR }  & \textbf{TNR }  \\ \thickhline
			\textbf{\begin{tabular}[c]{@{}c@{}} 6.7 \end{tabular}} & {1.0000} & 1.0000  & 0.9996  \\ \hline
			\textbf{\begin{tabular}[c]{@{}c@{}} 7.0 \end{tabular}} & {0.9998} & 0.9998  & 0.9989 \\ \hline
			\textbf{\begin{tabular}[c]{@{}c@{}} 7.3 \end{tabular}} & {0.9995} & 0.9996  & 0.9990 \\ \hline
			\textbf{\begin{tabular}[c]{@{}c@{}} 7.6 \end{tabular}} & {0.9995} & 0.9997  & 0.9988 \\ \hline
			\textbf{\begin{tabular}[c]{@{}c@{}} 7.9 \end{tabular}} & {0.9995} & 0.9996  & 0.9993  \\ \hline
	\end{tabular}}
	
\end{table}

Next, we investigate the performance of the classifier surrogate  in network connectivity prediction for a probabilistic earthquake event, i.e. for earthquakes with probabilistic magnitudes.  Following \cite{cosentino1977truncated,kang2008matrix}, it is assumed that the earthquake magnitude follows a truncated exponential distribution with the following pdf

\begin{equation} \label{magnitude_exponential}
f_{M}(m)=\left\{\begin{matrix}
\dfrac{\beta \exp\left [ -\beta (m-m_{\min}) \right ]}{1-\exp\left [ -\beta (m_{\max}-m_{\min}) \right ]}, & \; \; \; m_{\ell}\leqslant m\leqslant m_{u}, \\  
0, \; \; \; \; \; \; \; \; \; \; \; \; \; \; \; \; \; \; \; \; \; \; \; \; \; \; \; \; \; \; \; \; \; \; \; \; \: & \! \! \! \! \! \textup{otherwise},
\end{matrix}\right.
\end{equation}
where $m_{\min}$ and $m_{\max}$ are the minimum and maximum of random magnitudes,  which are set to 6.8 and 7.5, respectively. $\beta$ is the shape parameter and is set to 0.76 \cite{kang2008matrix} .

\begin{figure}
	\centering
	\includegraphics[width=0.6\linewidth]{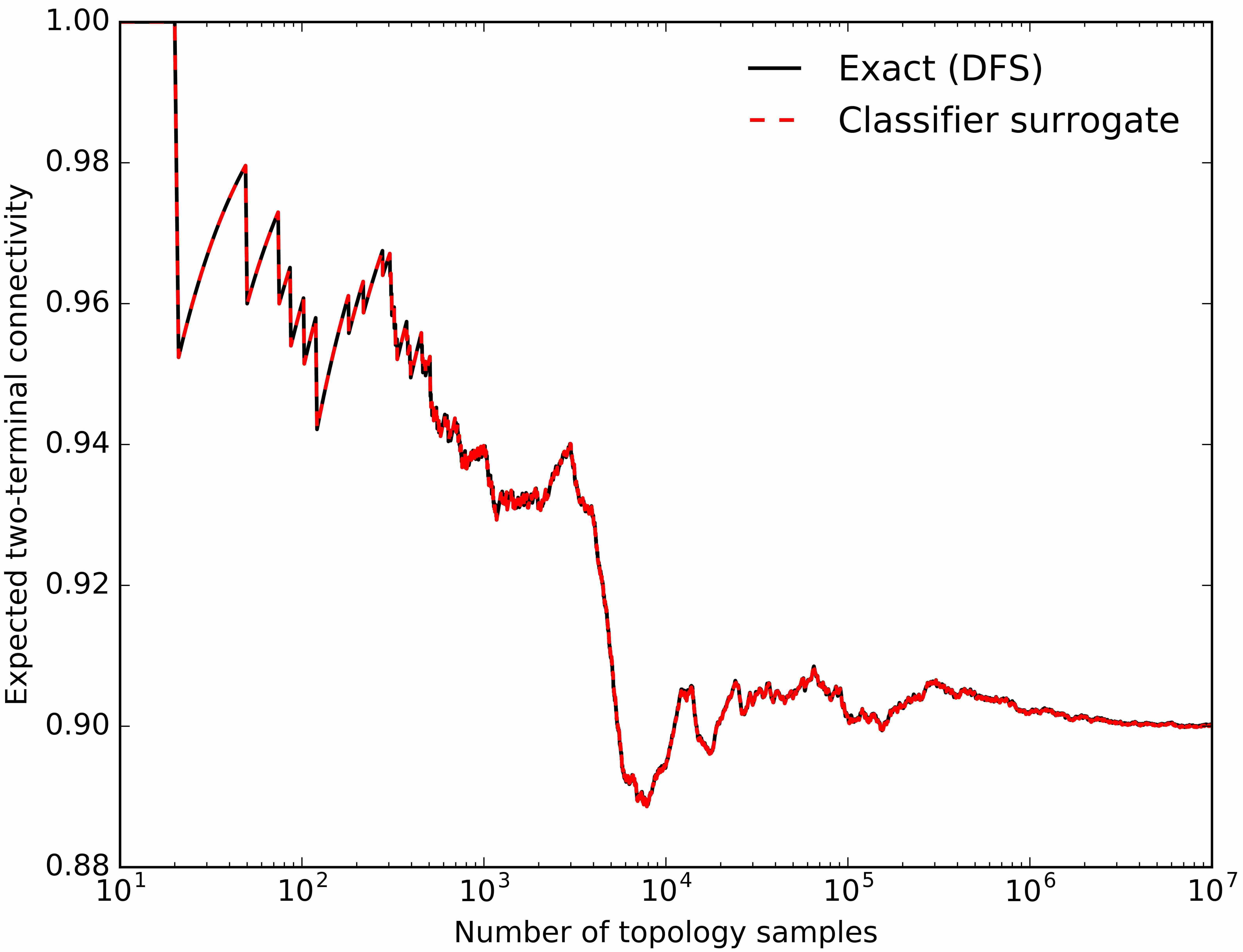}
	\captionsetup{}
	\caption{Convergence plot for prediction of two-terminal connectivity for a probabilistic earthquake event. The x-axis is in logarithmic scale.} 
	\label{fig.example1_convergence}
\end{figure}

A total of 10,000,000 network realizations are generated, each corresponding to a random sample from the probabilistic magnitude and a random sample from roadway failure states. For each of these network realizations, the two-terminal connectivity is evaluated using DFS algorithm and the classifier surrogate, and the resulting expected connectivities are compared in Figure \ref{fig.example1_convergence}. Expected two-terminal connectivity using DFS and classifier surrogate is, respectively, 0.9002 and 0.9001. The DFS and surrogate computational time are, respectively, 603.97 and 53.89 seconds, and $\mathbf{\alpha_{binary}}$, TPR, and TNR are 0.9997, 0.9998, and 0.9989, respectively. Once again, classifier surrogate results are in close agreement with DFS results with accuracies of more than 99.9\%, however, are achieved one order of magnitude faster.

\subsection{Uncertainty-Aware Two-Terminal Reliability Analysis Using Classifier Surrogate} \label{example2}

\begin{figure}
	\centering
	\includegraphics[width=0.6\linewidth]{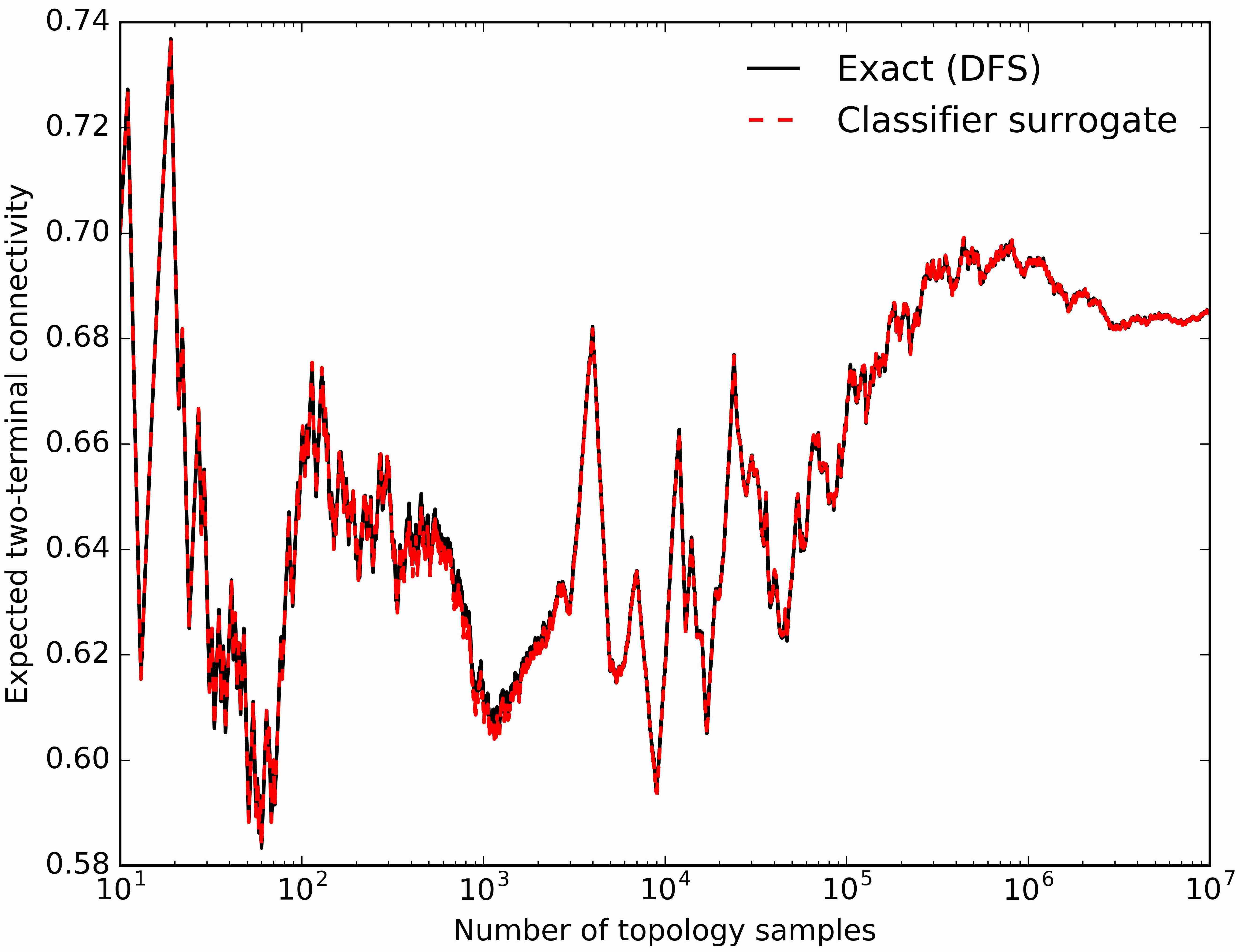}
	\captionsetup{}
	\caption{Convergence plot for prediction of two-terminal connectivity for a probabilistic earthquake event. The residual variability in GMPE is taken into account. The x-axis is in logarithmic scale.} 
	\label{fig.example2_convergence}
\end{figure}

In this section, we investigate an additional layer of uncertainty in two-terminal reliability assessment. Specifically, we consider the residual variability in GMPE, which is due to the model fitting error. To quantify the impact of this residual variability, we start off by  the probabilistic earthquake event defined in Equation \ref{magnitude_exponential}, and then consider  $a_\textup{PGA}$ and $S_a$ at each bridge location to be normally distributed random variables with mean values calculated using equations \ref{PGA} and \ref{Sa}, and standard deviations reported in \cite{graizer2016summary}. To study the classifier surrogate performance in this case, following the procedure represented in Figure \ref{flowchart1}, 10,000,000  network realizations are generated by consecutive sampling from earthquake magnitudes, $a_\textup{PGA}$ and $S_a$ at bridge locations, and roadway failure states according to their failure probabilities. Figure \ref{fig.example2_convergence} shows good agreement between expected two-terminal connectivities  using DFS and the classifier surrogate. This is while the surrogate evaluation, compared to DFS, is about one order of magnitude faster, i.e. 574.48 vs. 53.08 seconds. The estimated expected two-terminal connectivity using DFS and classifier surrogate are, respectively, 0.6853 and 0.6853, and $\mathbf{\alpha_{binary}}$, TPR, and TNR are 0.9990, 0.9993, and 0.9985, respectively.

It should be noted that the evaluated two-terminal reliability in this section (0.69) is smaller than the one evaluated in the previous section (0.90). This  highlights the importance of the additional uncertainties in ground motion intensity measures (e.g. $a_{\textup{PGA}}$ and $S_a$), which  is usually ignored in reliability studies (e.g. \cite{bocchini2011stochastic,stern2017accelerated}) and can lead to overestimation of the network reliability. 

\subsection{End-to-End Surrogate Training and Prediction} \label{example3}

As mentioned earlier,  two terminal connectivity calculations can be substantially accelerated by using an end-to-end surrogate, which  replaces the entire MCS as outlined in Figure \ref{flowchart2}. To numerically demonstrate this, we need to first train the end-to-end surrogate.  The training data can be generated using the DFS algorithm. Alternatively, we will use the previously-developed classifier surrogate to produce the training data. It should be noted that this training data set is not exact, but according to the results in the previous sections the error is expected to be negligible and the computational speed up is expected to be substantial. 

\begin{figure}
	\centering
	\includegraphics[width=1.0\linewidth]{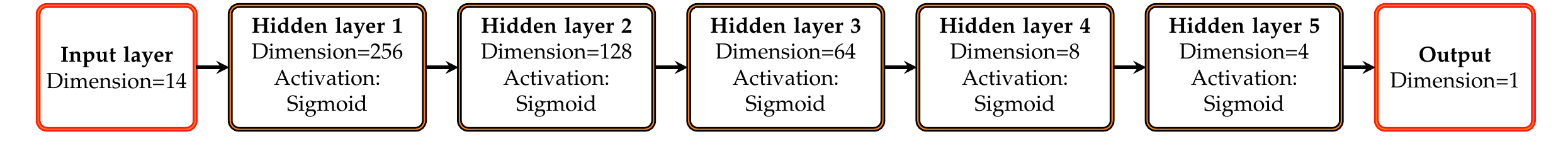}
	\captionsetup{}
	\caption{Architecture of the end-to-end surrogate. This surrogate is to be used instead of the MCS in order to accelerate the evaluation of expected two-terminal connectivity given the bridge failure probabilities. The model consists of 5 hidden layers with different dimensionalities. The input and output to this model are, respectively, a vector of roadway failure probabilities and a scalar that represents the expected two-terminal connectivity.} 
	\label{Architecture2}
\end{figure}

Figure \ref{Architecture2} shows the architecture of the end-to-end surrogate, with the input being the vector of roadway failure probabilities and the output the expected two-terminal connectivity. The surrogate consists of 5 hidden layers with different dimensionalities. Sigmoid activation is adopted for hidden layers. The Adam optimizer is used to minimize the Euclidean loss function (Equation \ref{MSE Loss}). To generate training and evaluation data, a total of 3,000 magnitude samples are drawn, and for each magnitude sample, 100,000 topology samples are drawn whose two-terminal connectivities were evaluated using the classifier surrogate. For a batch size of 64 and 2,000 epochs, the end-to-end surrogate training time (including generation of training and evaluation data  and model calibration) was 351.99 seconds.

\begin{figure}
	\centering
	\includegraphics[width=0.6\linewidth]{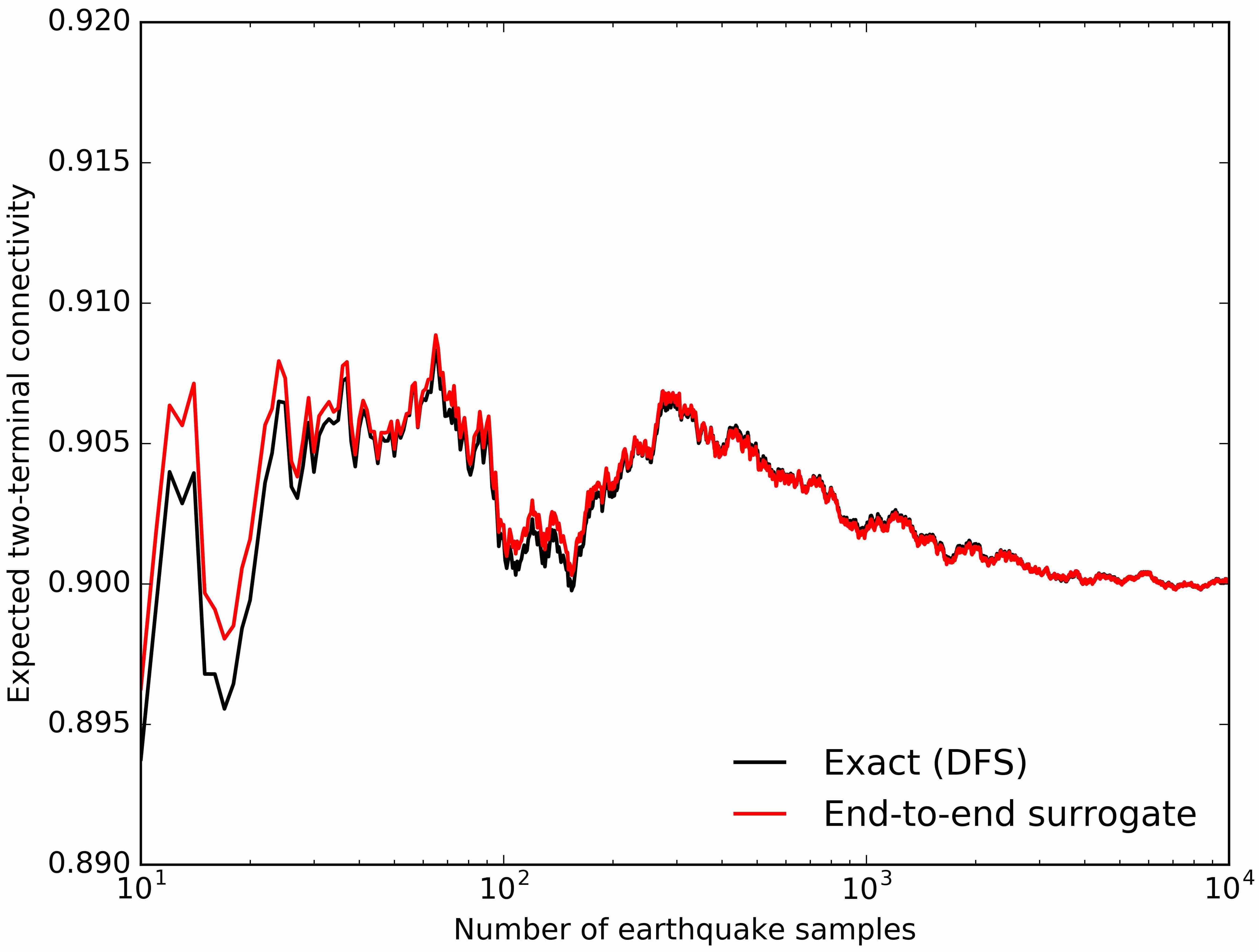}
	\captionsetup{}
	\caption{Convergence plot for prediction of two-terminal connectivity for a probabilistic earthquake event. The end-to-end surrogate estimates are calculated with no MCS.}
	\label{fig.example3_convergence}
\end{figure}

\begin{figure}
	\centering
	\includegraphics[width=0.6\linewidth]{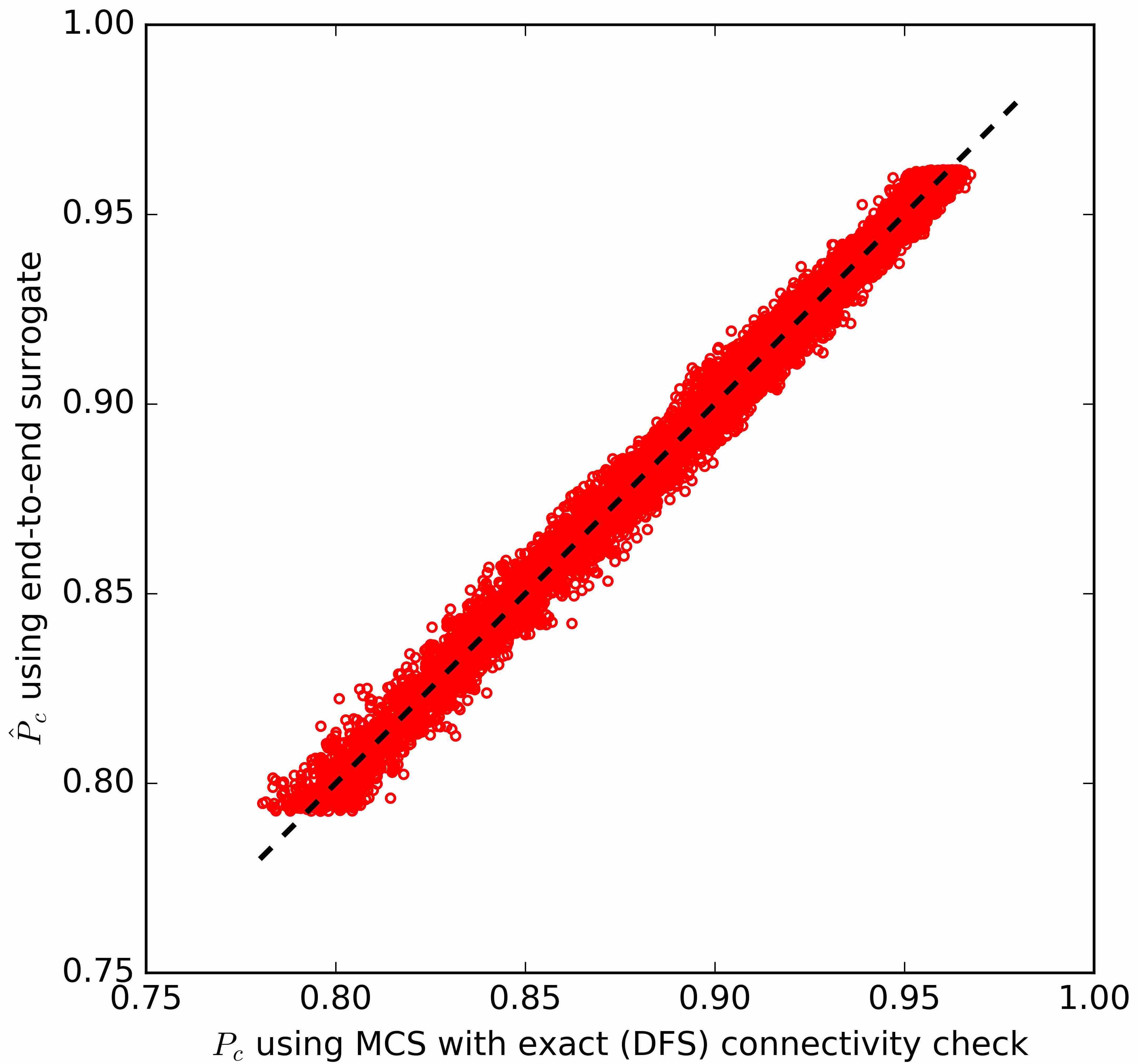}
	\captionsetup{}
	\caption{A comparison between the DFS and end-to-end surrogate predictions of the two-terminal connectivity for each earthquake realization. The surrogate predictions are calculated with no Monte Carlo simulation.}
	\label{fig.example3_individual}
\end{figure}

Using the trained end-to-end surrogate, we study the two-terminal connectivity of the San Jose-Mountain View transportation network subject to a probabilistic earthquake event. Similar to Section \ref{example1}, it is assumed that the earthquake magnitude follows a truncated exponential distribution. The lower and upper bounds for the magnitude variability are set to 6.8 and 7.5, respectively. Without loss of generality, the GMPE residual variabilities were ignored for simplicity.  To test the surrogate, 10,000 magnitude samples are drawn and for each sample, the expected two-terminal connectivity is calculated using  the end-to-end surrogate. As the reference case, for each earthquake realization, a total of 100,000 topology realizations are drawn and their connectivity is evaluated using DFS algorithm, and the results are compared in Figure \ref{fig.example3_convergence}. As another way of demonstrating the surrogate accuracy, Figure \ref{fig.example3_individual}  compares the DFS and surrogate predictions of connectivity for each earthquake realization. The expected two-terminal connectivity was estimated  to be 0.9001 using both approaches while the computational times for DFS and end-to-end surrogate were found to be 7,857.92 and 0.71 seconds, respectively. 

\subsection{One-at-a-time Sensitivity Analysis Using End-to-End Surrogate} \label{example4}
In this section, we demonstrate the application of the proposed end-to-end surrogate in maintenance planning. In particular, we consider the optimal seismic retrofitting of bridges \cite{buckle2006seismic} where decision makers  seek to improve the two-terminal reliability of the network. In this case, typically in the face of budget constraints, it is crucial to identify the  bridges that are most influential on two-terminal reliability and prioritize them for repair. To this end, a  one-at-a-time (OAT) sensitivity analysis can be performed \cite{komkov1986design}. It involves considering amplifications on the survival probabilities, one bridge at a time, while keeping the other bridges' survival probabilities at their nominal values. These amplifications should reflect the expected outcomes of repair plans for each bridge. We assume that these retrofit plans will result in an amplification rate of $10\%$ for every bridge.  Considering this rate, the expected two-terminal connectivities are then calculated using the DFS algorithm and end-to-end surrogate, for the nominal and ``retrofitted" networks. Here we consider a probabilistic earthquake event as defined in Equation \ref{magnitude_exponential} with a magnitude ranging between 7.3 and 7.9 $M_w$.  Table \ref{table: sensitivity analysis} shows the OAT sensitivity analysis results. For brevity, only the results for the three most and least sensitive components are shown. For the DFS results, for each earthquake realization, a total of 100,000 topology realizations are drawn. With no amplification, the expected two-terminal connectivity probability of the network subject to this probabilistic earthquake event is found to be 0.7641 using DFS and 0.7643 using the end-to-end surrogate. This table also highlights the substantial computational savings that the end-to-end surrogate can offer in repetitive processes, e.g. optimization, sensitivity analysis, or real-time risk-informed decision making.

\begin{table}[H]
	\centering
	\captionsetup{}
	\caption{Sensitivity analysis summary for selected bridges}
	\label{table: sensitivity analysis}
	\scalebox{0.9}{
			\begin{tabular}{|M{1.0cm}|M{1.0cm}|M{3.75cm}|M{3.75cm}|M{2.0cm}|M{2.0cm}|}
			\hline
			\textbf{Rank} & \textbf{Bridge ID} & \textbf{Improvement in connectivity (\%) (DFS estimate) }  &  \textbf{Improvement in connectivity (\%) (surrogate estimate)}  &
			\textbf{DFS time (s)}  & \textbf{Surrogate time (s) } \\ \thickhline
			\textbf{\begin{tabular}[c]{@{}c@{}} \end{tabular}} 1 & 0 & 8.22  & 8.19 & 8361.22 & 0.75 \\ \hline
			\textbf{\begin{tabular}[c]{@{}c@{}} \end{tabular}} 2 & 13 & 7.80  & 7.78& 7855.76 & 0.72 \\ \hline
			\textbf{\begin{tabular}[c]{@{}c@{}} \end{tabular}} 3 & 25 & {2.17} & 2.14 & 8334.18  & 0.74 \\ \hline
			\textbf{\begin{tabular}[c]{@{}c@{}} \end{tabular}} \textbf{\vdots} & \textbf{\vdots} & \textbf{\vdots} & \textbf{\vdots} & \textbf{\vdots} & \textbf{\vdots} \\ \hline
			\textbf{\begin{tabular}[c]{@{}c@{}} \end{tabular}} 37 & 14 & {0.07} & 0.06 & 8089.66 & 0.75 \\ \hline
			\textbf{\begin{tabular}[c]{@{}c@{}} \end{tabular}} 38 & 10 & {0.01} & 0.01 & 8131.76 & 0.74 \\ \hline
			\textbf{\begin{tabular}[c]{@{}c@{}} \end{tabular}} 39 & 26 & {0.00} & 0.00  & 8079.06 & 0.75 \\ \hline
	\end{tabular}}
	
\end{table}

\section{Conclusion}
Approximations and uncertainties inherent in infrastructure systems reliability analysis on one hand and the associated computational challenge on the other hand motivate the utilization of fast and sufficiently accurate surrogates that can replace one or more computational modules in the analysis pipeline. The resulting surrogate-based reliability analysis  can then facilitate optimal planning and management of infrastructure systems subject to natural hazards. In this paper, we studied the surrogates that are trained based on deep learning, and using a case study, highlighted how they can offer fast computation of infrastructure  response with high accuracy. An important advantage of using deep learning in building surrogates for nonlinear system responses is its capability for  automatic feature engineering/detection. This will  remove the need to manually identify  features for a given data, and make the approach broadly applicable to various nonlinear responses.

The proposed surrogate-based reliability analysis framework can be further extended by augmenting the training data to improve the prediction accuracy. An example of data augmentation for improving TNR can be created as follows. For each topology realization with no source-to-terminal connectivity, we can generate multiple additional topology realizations  by randomly (according to roadway failure probabilities) letting the survived roadways fail. These additional network realizations will not incur extra computational burden, as they are already known to be corresponding to a ``no-connectivity'' condition.
Another extension to further improve the computational efficiency is to make use of graphic processing units (GPUs) in deep neural network surrogate training and prediction. Deep learning generally involves large matrix multiplications that are substantially parallelizable using GPUs, leading to significant acceleration. 
\section*{Acknowledgement}
This work used the Extreme Science and Engineering Discovery Environment (XSEDE), which is supported by National Science Foundation grant number ACI-1053575.     

\section*{References}

\bibliography{bibfile}

\begin{thebibliography}{10}
\expandafter\ifx\csname url\endcsname\relax
  \def\url#1{\texttt{#1}}\fi
\expandafter\ifx\csname urlprefix\endcsname\relax\def\urlprefix{URL }\fi
\expandafter\ifx\csname href\endcsname\relax
  \def\href#1#2{#2} \def\path#1{#1}\fi

\bibitem{zacks2012introduction}
S.~Zacks, Introduction to reliability analysis: probability models and
  statistical methods, Springer Science \& Business Media, 2012.

\bibitem{godschalk1999natural}
D.~Godschalk, Natural hazard mitigation: Recasting disaster policy and
  planning, Island Press, 1999.

\bibitem{paton2003disaster}
D.~Paton, Disaster preparedness: a social-cognitive perspective, Disaster
  Prevention and Management: An International Journal 12~(3) (2003) 210--216.

\bibitem{perry2007natural}
M.~Perry, Natural disaster management planning: A study of logistics managers
  responding to the tsunami, International Journal of Physical Distribution \&
  Logistics Management 37~(5) (2007) 409--433.

\bibitem{adie2001holistic}
C.~E. Adie, et~al., Holistic disaster recovery: Ideas for building local
  sustainability after a natural disaster, DIANE Publishing, 2001.

\bibitem{wang2008integrated}
Y.-M. Wang, J.~Liu, T.~M. Elhag, An integrated ahp--dea methodology for bridge
  risk assessment, Computers \& industrial engineering 54~(3) (2008) 513--525.

\bibitem{stern2017accelerated}
R.~Stern, J.~Song, D.~Work, Accelerated monte carlo system reliability analysis
  through machine-learning-based surrogate models of network connectivity,
  Reliability Engineering \& System Safety 164 (2017) 1--9.

\bibitem{bocchini2011stochastic}
P.~Bocchini, D.~M. Frangopol, A stochastic computational framework for the
  joint transportation network fragility analysis and traffic flow distribution
  under extreme events, Probabilistic Engineering Mechanics 26~(2) (2011)
  182--193.

\bibitem{chang2010transportations}
L.~Chang, A.~S. Elnashai, B.~F. Spencer, J.~Song, Y.~Ouyang, Transportations
  systems modeling and applications in earthquake engineering, Tech. rep., DTIC
  Document (2010).

\bibitem{bocchini2011generalized}
P.~Bocchini, D.~M. Frangopol, Generalized bridge network performance analysis
  with correlation and time-variant reliability, Structural Safety 33~(2)
  (2011) 155--164.

\bibitem{nabian2017uncertainty}
M.~A. Nabian, H.~Meidani, Uncertainty quantification and pca-based model
  reduction for parallel monte carlo analysis of infrastructure system
  reliability, Tech. rep. (2017).

\bibitem{liu2009two}
C.~Liu, Y.~Fan, F.~Ord{\'o}{\~n}ez, A two-stage stochastic programming model
  for transportation network protection, Computers \& Operations Research
  36~(5) (2009) 1582--1590.

\bibitem{faturechi2014measuring}
R.~Faturechi, E.~Miller-Hooks, Measuring the performance of transportation
  infrastructure systems in disasters: A comprehensive review, Journal of
  infrastructure systems 21~(1) (2014) 04014025.

\bibitem{koziel2013surrogate}
S.~Koziel, L.~Leifsson, Surrogate-based modeling and optimization, Applications
  in Engineering.

\bibitem{queipo2005surrogate}
N.~V. Queipo, R.~T. Haftka, W.~Shyy, T.~Goel, R.~Vaidyanathan, P.~K. Tucker,
  Surrogate-based analysis and optimization, Progress in aerospace sciences
  41~(1) (2005) 1--28.

\bibitem{wild2008orbit}
S.~M. Wild, R.~G. Regis, C.~A. Shoemaker, Orbit: Optimization by radial basis
  function interpolation in trust-regions, SIAM Journal on Scientific Computing
  30~(6) (2008) 3197--3219.

\bibitem{kleijnen2009kriging}
J.~P. Kleijnen, Kriging metamodeling in simulation: A review, European journal
  of operational research 192~(3) (2009) 707--716.

\bibitem{tabatabaee2012two}
N.~Tabatabaee, M.~Ziyadi, Y.~Shafahi, Two-stage support vector classifier and
  recurrent neural network predictor for pavement performance modeling, Journal
  of Infrastructure Systems 19~(3) (2012) 266--274.

\bibitem{ziyadi2016efficient}
M.~Ziyadi, I.~L. Al-Qadi, Efficient surrogate method for predicting pavement
  response to various tire configurations, Neural Computing and Applications
  (2016) 1--13.

\bibitem{hornik1989multilayer}
K.~Hornik, M.~Stinchcombe, H.~White, Multilayer feedforward networks are
  universal approximators, Neural networks 2~(5) (1989) 359--366.

\bibitem{hornik1991approximation}
K.~Hornik, Approximation capabilities of multilayer feedforward networks,
  Neural networks 4~(2) (1991) 251--257.

\bibitem{lecun2015deep}
Y.~LeCun, Y.~Bengio, G.~Hinton, Deep learning, Nature 521~(7553) (2015)
  436--444.

\bibitem{schmidhuber2015deep}
J.~Schmidhuber, Deep learning in neural networks: An overview, Neural networks
  61 (2015) 85--117.

\bibitem{goodfellow2016deep}
I.~Goodfellow, Y.~Bengio, A.~Courville, Deep learning, MIT Press, 2016.

\bibitem{demuth2014neural}
H.~B. Demuth, M.~H. Beale, O.~De~Jess, M.~T. Hagan, Neural network design,
  Martin Hagan, 2014.

\bibitem{cheng2008new}
J.~Cheng, Q.~Li, R.-c. Xiao, A new artificial neural network-based response
  surface method for structural reliability analysis, Probabilistic Engineering
  Mechanics 23~(1) (2008) 51--63.

\bibitem{papadrakakis2002reliability}
M.~Papadrakakis, N.~D. Lagaros, Reliability-based structural optimization using
  neural networks and monte carlo simulation, Computer methods in applied
  mechanics and engineering 191~(32) (2002) 3491--3507.

\bibitem{gomes2004comparison}
H.~M. Gomes, A.~M. Awruch, Comparison of response surface and neural network
  with other methods for structural reliability analysis, Structural safety
  26~(1) (2004) 49--67.

\bibitem{hurtado2001neural}
J.~E. Hurtado, D.~A. Alvarez, Neural-network-based reliability analysis: a
  comparative study, Computer methods in applied mechanics and engineering
  191~(1) (2001) 113--132.

\bibitem{srivaree2002estimation}
C.~Srivaree-ratana, A.~Konak, A.~E. Smith, Estimation of all-terminal network
  reliability using an artificial neural network, Computers \& Operations
  Research 29~(7) (2002) 849--868.

\bibitem{elhewy2006reliability}
A.~H. Elhewy, E.~Mesbahi, Y.~Pu, Reliability analysis of structures using
  neural network method, Probabilistic Engineering Mechanics 21~(1) (2006)
  44--53.

\bibitem{cardoso2008structural}
J.~B. Cardoso, J.~R. de~Almeida, J.~M. Dias, P.~G. Coelho, Structural
  reliability analysis using monte carlo simulation and neural networks,
  Advances in Engineering Software 39~(6) (2008) 505--513.

\bibitem{zhang2004performance}
J.~Zhang, R.~O. Foschi, Performance-based design and seismic reliability
  analysis using designed experiments and neural networks, Probabilistic
  Engineering Mechanics 19~(3) (2004) 259--267.

\bibitem{kang2008matrix}
W.-H. Kang, J.~Song, P.~Gardoni, Matrix-based system reliability method and
  applications to bridge networks, Reliability Engineering \& System Safety
  93~(11) (2008) 1584--1593.

\bibitem{stewart2015selection}
J.~P. Stewart, J.~Douglas, M.~Javanbarg, Y.~Bozorgnia, N.~A. Abrahamson, D.~M.
  Boore, K.~W. Campbell, E.~Delavaud, M.~Erdik, P.~J. Stafford, Selection of
  ground motion prediction equations for the global earthquake model,
  Earthquake Spectra 31~(1) (2015) 19--45.

\bibitem{bommer2010selection}
J.~J. Bommer, J.~Douglas, F.~Scherbaum, F.~Cotton, H.~Bungum, D.~F{\"a}h, On
  the selection of ground-motion prediction equations for seismic hazard
  analysis, Seismological Research Letters 81~(5) (2010) 783--793.

\bibitem{douglas2017ground}
J.~Douglas, Ground-motion prediction equations 1964-2016, Pacific Earthquake
  Engineering Research Center Berkeley, CA, 2017.

\bibitem{graizer2016summary}
V.~Graizer, E.~Kalkan, Summary of the gk15 ground-motion prediction equation
  for horizontal pga and 5\% damped psa from shallow crustal continental
  earthquakes, Bulletin of the Seismological Society of America 106~(2) (2016)
  687--707.

\bibitem{graizer2015update}
V.~Graizer, Update of the graizer--kalkan groundmotion prediction equations for
  shallow crustal continental earthquakes, US Geol. Surv. Open-File Rept 1009
  (2015) 98.

\bibitem{fema2008hazus}
FEMA, Hazus-mh mr3: Technical manual (2008).

\bibitem{tarjan1972depth}
R.~Tarjan, Depth-first search and linear graph algorithms, SIAM journal on
  computing 1~(2) (1972) 146--160.

\bibitem{korf1985depth}
R.~E. Korf, Depth-first iterative-deepening: An optimal admissible tree search,
  Artificial intelligence 27~(1) (1985) 97--109.

\bibitem{rosenthal2000parallel}
J.~S. Rosenthal, Parallel computing and monte carlo algorithms, Far east
  journal of theoretical statistics 4~(2) (2000) 207--236.

\bibitem{jin2000improvements}
W.~Jin, Z.~J. Li, L.~S. Wei, H.~Zhen, The improvements of bp neural network
  learning algorithm, in: Signal Processing Proceedings, 2000. WCCC-ICSP 2000.
  5th International Conference on, Vol.~3, IEEE, 2000, pp. 1647--1649.

\bibitem{baldi2000assessing}
P.~Baldi, S.~Brunak, Y.~Chauvin, C.~A. Andersen, H.~Nielsen, Assessing the
  accuracy of prediction algorithms for classification: an overview,
  Bioinformatics 16~(5) (2000) 412--424.

\bibitem{schult2008exploring}
D.~A. Schult, P.~Swart, Exploring network structure, dynamics, and function
  using networkx, in: Proceedings of the 7th Python in Science Conferences
  (SciPy 2008), Vol. 2008, 2008, pp. 11--16.

\bibitem{chollet2015keras}
F.~Chollet, Keras, \url{https://github.com/fchollet/keras} (2015).

\bibitem{mohammad_amin_nabian_2017_846898}
M.~A. Nabian,
  \href{https://doi.org/10.5281/zenodo.846898}{{UIUC-UQ/Deep-Learning-for-Reliability-Analysis:
  Deep Learning for Accelerated Reliability Analysis of Infrastructure
  Systems}} (Aug. 2017).
\newblock \href {http://dx.doi.org/10.5281/zenodo.846898}
  {\path{doi:10.5281/zenodo.846898}}.
\newline\urlprefix\url{https://doi.org/10.5281/zenodo.846898}

\bibitem{kanamori1977energy}
H.~Kanamori, The energy release in great earthquakes, Journal of geophysical
  research 82~(20) (1977) 2981--2987.

\bibitem{kingma2014adam}
D.~Kingma, J.~Ba, Adam: A method for stochastic optimization, arXiv preprint
  arXiv:1412.6980.

\bibitem{cosentino1977truncated}
P.~Cosentino, V.~Ficarra, D.~Luzio, Truncated exponential frequency-magnitude
  relationship in earthquake statistics, Bulletin of the Seismological Society
  of America 67~(6) (1977) 1615--1623.

\bibitem{buckle2006seismic}
G.~Buckle, I.~Friedland, J.~Mander, G.~Martin, R.~Nutt, M.~Power, Seismic
  retrofitting manual for highway structures, MCEER, Buffalo, 2006.

\bibitem{komkov1986design}
V.~Komkov, K.~K. Choi, E.~J. Haug, Design sensitivity analysis of structural
  systems, Vol. 177, Academic press, 1986.

\end{thebibliography}

\end{document}